\documentclass[twocolumn,english,twoside]{IEEEtran}
\usepackage[T1]{fontenc}
\usepackage{xcolor}
\usepackage{float}
\usepackage{amsthm}
\usepackage{amsmath}
\usepackage{amssymb}
\usepackage{graphicx}
\PassOptionsToPackage{normalem}{ulem}
\usepackage{ulem}

\makeatletter

\floatstyle{ruled}
\newfloat{algorithm}{tbp}{loa}
\providecommand{\algorithmname}{Algorithm}
\floatname{algorithm}{\protect\algorithmname}
\providecolor{lyxadded}{rgb}{0,0,1}
\providecolor{lyxdeleted}{rgb}{1,0,0}
\theoremstyle{plain}
\newtheorem{thm}{\protect\theoremname}
\theoremstyle{plain}
\newtheorem{lem}[thm]{\protect\lemmaname}

\usepackage{amsmath}
\usepackage[numbers,sort]{natbib}

\usepackage{graphicx}
\usepackage{subfigure}

\usepackage{pdfsync}

\makeatother

\usepackage{babel}
\providecommand{\lemmaname}{Lemma}
\providecommand{\theoremname}{Theorem}

\begin{document}

\title{\vspace{-0.6cm}A Parallel Stochastic Approximation Method for Nonconvex
Multi-Agent Optimization Problems}

\author{Yang Yang, Gesualdo Scutari, Daniel P. Palomar, and Marius Pesavento\vspace{-0.3cm}%
\thanks{{\footnotesize{}Y. Yang and M. Pesavento are with Communication Systems
Group, Darmstadt University of Technology, Darmstadt, Germany. G.
Scutari is with the Department of Electrical Engineering, State University
of New York at Buffalo, Buffalo, USA. D. P. Palomar is with the Department
of} {\footnotesize{}Electronic and Computer Engineering, Hong Kong
University of Science and Technology, Hong Kong. Emails: }\texttt{\footnotesize{}<yang,pesavento>@nt.tu-darmstadt.de}{\footnotesize{};
}\texttt{\footnotesize{}gesualdo@buffalo.edu}{\footnotesize{}; }\texttt{\footnotesize{}palomar@ust.hk}{\footnotesize{}.}{\footnotesize \par}

Part of this work has been presented at the 14th IEEE Workshop on
Signal Processing Advances in Wireless Communications (SPAWC), Jun.
2013 \cite{Yang2013}.%
}\vspace{-0.4cm}}
\maketitle
\begin{abstract}
Consider the problem of minimizing the expected value of a (possibly\emph{
nonconvex}) cost function parameterized by a random (vector) variable,
when the expectation cannot be computed accurately (e.g., because
the statistics of the random variables are unknown and/or the computational
complexity is prohibitive). Classical sample stochastic gradient methods
for solving this problem may empirically suffer from slow convergence.\textcolor{blue}{{}
}In this paper, we propose for the first time a stochastic \emph{parallel}
\emph{Successive Convex Approximation}-based\emph{ }(best-response)\emph{
}algorithmic framework for general \emph{nonconvex }stochastic sum-utility
optimization problems, which arise naturally in the design of multi-agent
systems. The proposed novel decomposition enables all users to update
their optimization variables \emph{in parallel} by solving a sequence
of strongly convex subproblems, one for each user. Almost surely convergence
to stationary points is proved. We then customize our algorithmic
framework to solve the stochastic sum rate maximization problem over
Single-Input-Single-Output (SISO) frequency-selective interference
channels, multiple-input-multiple-output (MIMO) interference channels,
and MIMO multiple-access channels. Numerical results show that our
algorithms are much faster than state-of-the-art stochastic gradient
schemes while achieving the same (or better) sum-rates.\end{abstract}

\begin{IEEEkeywords}
Multi-agent systems, parallel optimization, stochastic approximation.\vspace{-0.3cm}
\end{IEEEkeywords}

\section{Introduction}

Wireless networks are composed of users that may have different objectives
and generate interference when no multiplexing scheme is imposed to
regulate the transmissions; examples are peer-to-peer networks, cognitive
radio systems, and ad-hoc networks. A usual and convenient way of
designing such multi-user systems is by optimizing the ``social function'',
i.e., the (weighted) sum of the users' objective functions. This formulation
however requires the knowledge of the system parameters, which in
practice is either difficult to acquire (e.g., when the parameters
are rapidly changing) or imperfect due to estimation errors. In such
scenarios, it is convenient to focus on the optimization of long-term
performance of the system, measured as the expected value of the social
function parametrized by the random system parameters. In this paper,
we consider the frequent and difficult case wherein (the expected
value of) the social function is \emph{nonconvex} and the expectation
cannot be computed (either numerically or in closed form). Such a
system design naturally falls into the class of stochastic optimization
\cite{Robbins1951,KushnerYin2003}.

Gradient methods for \emph{unconstrained} stochastic \emph{nonconvex}
optimization problems have been studied in \cite{polyak1987introduction,Bertsekas2000,Tsitsiklis1986},
where almost sure convergence to stationary points has been established,
under some technical conditions; see, e.g., \cite{Bertsekas2000}.
The extension of these methods to \emph{constrained} optimization
problems is not straightforward; in fact, the descent-based convergence
analysis developed for unconstrained gradient methods no longer applies
to their projected counterpart (due to the presence of the projection
operator). Convergence of stochastic gradient \emph{projection} methods
has been proved only for \emph{convex} objective functions \cite{Ermol'ev1972,polyak1987introduction,Yousefian2012}.

To cope with nonconvexity, \emph{gradient averaging} seems to be an
essential step to resemble convergence; indeed, stochastic \emph{conditional}
gradient methods for \emph{nonconvex} constrained problems hinge on
this idea \cite{Ermol'ev1977,Ruszczynski1980,Ruszczynski2008,Nemirovski2009}:
at each iteration the new update of the variables is based on the
average of the current and past gradient samples. Under some technical
conditions, the average sample gradient eventually resembles the nominal
(but unavailable) gradient of the (stochastic) objective function
\cite{Ermol'ev1977,Gupal1974a}; convergence analysis can then borrow
results from deterministic nonlinear programming. 

Numerical experiments for large classes of problems show that plain
gradient-like methods usually converge slowly and are very sensitive
to the choice of the step-size. Some acceleration techniques have
been proposed in the literature \cite{Polyak1992,Yousefian2012},
but only for\emph{ strongly convex} objective functions. Here we are
interested in nonconvex (constrained) stochastic problems. Moreover,
(proximal, accelerated) stochastic gradient-based schemes use only
the first order information of the objective function (or its realizations);
recently it was shown \cite{Scutarib,Scutari_BigData,Scutari_hybrid}
that for deterministic nonconvex optimization problems exploiting
the structure of the function by replacing its linearization with
a \textquotedblleft better\textquotedblright{} approximant can enhance
empirical convergence speed. In this paper we aim at bringing this
idea into the context of stochastic optimization problems. 

Our main contribution is to develop a new, broad algorithmic framework
for the computation of stationary solutions of a wide class of (stochastic)
nonconvex optimization problems, encompassing many multi-agent system
designs of practical interest. The essential idea underlying our approach
is to decompose the original nonconvex \emph{stochastic} problem into
a sequence of (simpler) deterministic subproblems whereby the objective
function is replaced by suitable chosen \emph{sample convex} approximations;
the subproblems can be then solved in a \emph{parallel} and \emph{distributed}
fashion across the users. Other key features of our framework are:
i) no knowledge of the objective function parameters (e.g., the Lipschitz
constant of the gradient) is required; ii) it is very flexible in
the choice of the approximant of the nonconvex objective function,
which need not be necessarily its first or second order approximation
(like in proximal-gradient schemes); of course it includes, among
others, updates based on stochastic gradient- or Newton-type approximations;
iii) it can be successfully used also to \emph{robustify} distributed
iterative algorithms solving \emph{deterministic} social problems,
when only inexact estimates of the system parameters are available;
and iv) it encompasses a gamut of novel algorithms, offering a wide
flexibility to control iteration complexity, communication overhead,
and convergence speed, \emph{while converging under the same conditions}.
These desirable features make our schemes applicable to several different
problems and scenarios. As illustrative examples, we customize our
algorithms to some resource allocations problems in wireless communications,
namely: the sum-rate maximization problems over MIMO Interference
Channels (ICs) and Multiple Access Channels (MACs). The resulting
algorithms outperform existing (gradient-based) methods both theoretically
and numerically. 

The proposed decomposition technique hinges on successive convex approximation
(SCA) methods, and it is a nontrivial generalization of \cite{Scutarib}
to stochastic optimization problems. An SCA framework for stochastic
optimization problems has also been proposed in a recent submission
\cite{Razaviyayn_2013}%
\footnote{A preliminary version of our work appeared independently before \cite{Razaviyayn_2013}
at IEEE SPAWC 2013 \cite{Yang2013}.%
}; however our method differs from \cite{Razaviyayn_2013} in many
features. First of all, we relax the key requirement that the convex
approximation must be a tight global upper bound of the (sample) objective
function, as required instead in \cite{Razaviyayn_2013}. This represents
a turning point in the design of distributed stochastic SCA-based
methods, enlarging substantially the class of (large scale) stochastic
nonconvex problems solvable using our framework. Second, even when
the aforementioned constraint can be met, it is not always guaranteed
that the resulting convex (sample) subproblems are decomposable across
the users, implying that a centralized implementation might be required
in \cite{Razaviyayn_2013}; our schemes instead naturally lead to
a parallel and distributed implementation. Third, the proposed methods
converge under weaker conditions than those in \cite{Razaviyayn_2013}. 

Finally, within the classes of approximation-based methods for stochastic
optimization problems, it is worth mentioning the so-called Sample
Average Approach (SAA) \cite{Robinson1996,Linderoth2006,Mairal2010,Razaviyayn_2013}:
the ``true'' (stochastic) objective function is approximated by
an ensemble average. Then the resulting deterministic SSA optimization
problem has to be solved by an appropriate numerical procedure. When
the original objective function is nonconvex, however, computing the
global optimal solution of the SAA at each step may not be easy, if
not impossible. Therefore SSA-based methods are generally used to
solve stochastic \emph{convex} optimization problems only. 

The rest of the paper is organized as follows. Sec. \ref{sec:Problem-Formulation}
formulates the problem along with some interesting applications. The
novel stochastic decomposition framework is introduced in Sec. \ref{sec:The-Stochastic-Distributed};
customizations of the main algorithms to sample applications are discussed
in Sec. \ref{sec:applications}. Finally, Sec. \ref{sec:Conclusions}
draws some conclusions.\vspace{-0.2cm}

\section{\label{sec:Problem-Formulation}Problem Formulation}

We consider the design of a multi-agent system composed of $I$ users;
each user $i$ has his own strategy vector $\mathbf{x}_{i}$ to optimize,
which belongs to the feasible convex set $\mathcal{X}_{i}\subseteq\mathbb{C}^{n_{i}}$.
The variables of the other users are denoted by $\mathbf{x}_{-i}\triangleq\left(\mathbf{x}_{j}\right)_{j=1,j\neq i}^{I}$,
and the joint strategy set of all users is $\mathcal{X}=\mathcal{X}_{1}\times\ldots\times\mathcal{X}_{I}$. 

The stochastic social optimization problem is formulated as:\vspace{-0.25em}
\begin{equation}
\begin{array}{cl}
\underset{\mathbf{x}\triangleq\left\{ \mathbf{x}_{i}\right\} }{\textrm{minimize}} & U(\mathbf{x})\triangleq\mathbb{E}\Biggl[\:{\displaystyle {\sum_{j\in\mathcal{I}_{f}}}}\, f_{j}(\mathbf{x},\boldsymbol{\xi})\Biggr]\\
\textrm{subject to} & \mathbf{x}_{i}\in\mathcal{X}_{i},\quad i=1,\ldots,I,
\end{array}\label{eq:social-formulation}
\end{equation}
where $\mathcal{I}_{f}\triangleq\{1,\ldots,I_{f}\}$, with $I_{f}$
being the number of functions; each cost function $f_{j}(\mathbf{x},\boldsymbol{\xi}):\mathcal{X}\times\mathcal{D}\rightarrow\mathbb{R}$
depends on the joint strategy vector $\mathbf{x}$ and a random vector
$\boldsymbol{\xi}$, whose  probability distribution is defined on
a set $\mathcal{D}\subseteq\mathbb{C}^{m}$; and the expectation is
taken with respect to (w.r.t.)  $\boldsymbol{\xi}$. Note that the
optimization variables can be complex-valued; in such a case, all
the gradients of real-valued functions are intended to be conjugate
gradients \cite{Scutari2012a,Hjorungnes}.\smallskip

\noindent \textbf{Assumptions: }We make the following blanket assumptions:
\begin{itemize}
\item [(a)] Each $\mathcal{X}_{i}$ is compact and convex;
\item [(b)] Each $f_{j}(\bullet,\boldsymbol{\xi})$ is continuously differentiable
on $\mathcal{X}$, for any given $\boldsymbol{\xi}$, and the gradient
is Lipschitz continuous with constant $L_{\nabla f_{j}(\boldsymbol{\xi})}$.
Furthermore, the gradient of $U(\mathbf{x})$ is Lipschitz continuous
with constant $L_{\nabla U}<+\infty$.
\end{itemize}
These assumptions are quite standard and are satisfied by a large
class of problems. Note that the existence of a solution to (\ref{eq:social-formulation})
is guaranteed by Assumption (a). Since $U(\mathbf{x})$ is not assumed
to be jointly convex in $\mathbf{x}$, (\ref{eq:social-formulation})
is generally nonconvex. Some instances of (\ref{eq:social-formulation})
satisfying the above assumptions are briefly listed next.\smallskip

\noindent\emph{Example \#1: }Consider the maximization of the ergodic
sum-rate over frequency-selective ICs:
\begin{equation}
\begin{array}{cl}
\underset{\mathbf{p}_{1},\ldots,\mathbf{p}_{I}}{\textrm{maximize}} & \mathbb{E}\left[{\displaystyle {\sum_{n=1}^{N}\sum_{i=1}^{I}}}\log\left(1+\frac{|h_{ii,n}|^{2}p_{i,n}}{\sigma_{i,n}^{2}+\sum_{j\neq i}|h_{ij,n}|^{2}p_{j,n}}\right)\right]\medskip\\
\textrm{subject to} & \mathbf{p}_{i}\in\mathcal{P}_{i}\triangleq\{\mathbf{p}_{i}:\mathbf{p}_{i}\geq\mathbf{0},\mathbf{1}^{T}\mathbf{p}_{i}\leq P_{i}\},\,\forall i,
\end{array}\label{eq:SISO-IC-SR}
\end{equation}
where $\mathbf{p}_{i}\triangleq\left\{ p_{i,n}\right\} _{n=1}^{N}$
with $p_{i,n}$ being the transmit power of user $i$ on subchannel
(subcarrier) $n$, $N$ is the number of parallel subchannels, $P_{i}$
is the total power budget, $h_{ij,n}$ is the channel coefficient
from transmitter $j$ to receiver $i$ on subchannel $n$, and $\sigma_{i,n}^{2}$
is the variance of the thermal noise over subchannel $n$ at the receiver
$i$. The expectation is over channel coefficients $(h_{ij,n})_{i,j,n}$.
\smallskip

\noindent\emph{Example \#2: }The following maximization of the ergodic
sum-rate over MIMO ICs also falls into the class of problems (\ref{eq:social-formulation}):
\begin{equation}
\begin{array}{cl}
\underset{\mathbf{Q}_{1},\ldots,\mathbf{Q}_{I}}{\textrm{maximize}} & \mathbb{E}\left[{\displaystyle {\sum_{i=1}^{I}}}\log\det\left(\mathbf{I}+\mathbf{H}_{ii}\mathbf{Q}_{i}\mathbf{H}_{ii}^{H}\mathbf{R}_{i}(\mathbf{Q}_{-i},\mathbf{H})^{-1}\right)\right]\medskip\\
\textrm{subject to} & \mathbf{Q}_{i}\in\mathcal{Q}_{i}\triangleq\{\mathbf{Q}_{i}:\mathbf{Q}_{i}\succeq\mathbf{0},\textrm{Tr}(\mathbf{Q}_{i})\leq P_{i}\},\,\forall i,
\end{array}\label{eq:MIMO-IC-SR}
\end{equation}
where $\mathbf{R}_{i}\left(\mathbf{Q}_{-i},\mathbf{H}\right)\triangleq\mathbf{R}_{\textrm{N}_{i}}+\sum_{j\neq i}\mathbf{H}_{ij}\mathbf{Q}_{j}\mathbf{H}_{ij}^{H}$
is the covariance matrix of the thermal noise $\mathbf{R}_{\textrm{N}_{i}}$
(assumed to be full rank) plus the multi-user interference, $P_{i}$
is the total power budget, and the expectation in (\ref{eq:MIMO-IC-SR})
is taken over the channels $\mathbf{H}\triangleq(\mathbf{H}_{ij})_{i,j=1}^{I}$.\smallskip

\noindent\emph{Example \#3: }Another application of interest is the
maximization of the ergodic sum-rate over MIMO MACs:
\begin{equation}
\begin{array}{cl}
\underset{\mathbf{Q}_{1},\ldots,\mathbf{Q}_{I}}{\textrm{maximize}} & \mathbb{E}\left[\log\det\left(\mathbf{R}_{\textrm{N}}+\sum_{i=1}^{I}\mathbf{H}_{i}\mathbf{Q}_{i}\mathbf{H}_{i}^{H}\right)\right]\\
\textrm{subject to} & \mathbf{Q}_{i}\in\mathcal{Q}_{i},\,\forall i.
\end{array}\label{eq:MIMO-MAC-SR}
\end{equation}
This is a special case of (\ref{eq:social-formulation}) where the
utility function is concave in $\mathbf{Q}\triangleq(\mathbf{Q}_{i})_{i=1}^{I}$,
$I_{f}=1$, $\mathcal{I}_{f}=\{1\}$, and the expectation in (\ref{eq:MIMO-MAC-SR})
is taken over the channels $\mathbf{H}\triangleq(\mathbf{H}_{i})_{i=1}^{I}$.\smallskip

\noindent\emph{Example \#4: }The algorithmic framework that will
be introduced shortly can be successfully used also to \emph{robustify}
distributed iterative algorithms solving \emph{deterministic} (nonconvex)
social problems, but in the presence of \emph{inexact estimates} of
the system parameters. More specifically, consider as example the
following sum-cost minimization multi-agent problem:
\begin{equation}
\begin{array}{cl}
\underset{\mathbf{x}}{\textrm{minimize}} & \sum_{i=1}^{I}f_{i}(\mathbf{x}_{1},\ldots,\mathbf{x}_{I})\\
\textrm{subject to} & \mathbf{x}_{i}\in\mathcal{X}_{i},\quad i=1,\ldots,I,
\end{array}\label{eq:deterministic_social}
\end{equation}
where $f_{i}(\mathbf{x}_{i},\mathbf{x}_{-i})$ is uniformly convex
in $\mathbf{x}_{i}\in\mathcal{X}_{i}$. An efficient distributed algorithm
converging to stationary solutions of (\ref{eq:deterministic_social})
has been recently proposed in \cite{Scutarib}: at each iteration
$t$, given the current iterate $\mathbf{x}^{t}$, every agent $i$
minimizes (w.r.t. $\mathbf{x}_{i}\in\mathcal{X}_{i}$) the following
convexified version of the social function: 
\[
f_{i}(\mathbf{x}_{i},\mathbf{x}_{-i}^{t})+\bigl\langle\mathbf{x}_{i}-\mathbf{x}_{i}^{t},{\textstyle \sum_{j\neq i}}\nabla_{i}f_{j}(\mathbf{x}^{t})\bigr\rangle+\tau_{i}\left\Vert \mathbf{x}_{i}-\mathbf{x}_{i}^{t}\right\Vert ^{2},
\]
where $\nabla_{i}f_{j}(\mathbf{x})$ stands for $\nabla_{\mathbf{x}_{i}^{*}}f_{j}(\mathbf{x})$,
and $\left<\mathbf{a},\mathbf{b}\right>\triangleq\Re\left(\mathbf{a}^{H}\mathbf{b}\right)$
($\left\Vert \mathbf{a}\right\Vert =\sqrt{\left<\mathbf{a,a}\right>}$).
The evaluation of the above function requires the exact knowledge
of $\nabla_{i}f_{j}(\mathbf{x}^{t})$ for all $j\neq i$. In practice,
however, only a noisy estimate of $\nabla_{i}f_{j}(\mathbf{x}^{t})$
is available \cite{Zhang2008b,Hong2011,DiLorenzo2013}. In such cases,
convergence of pricing-based algorithms \cite{Huang2006a,Shi2009a,Kim2011,Scutarib}
is in jeopardy. We will show in Sec. IV-C that the proposed framework
can be readily applied, for example, to robustify (and make convergent),
e.g., pricing-based schemes, such as \cite{Huang2006a,Shi2009a,Kim2011,Scutarib}.

Since the class of problems (\ref{eq:social-formulation}) is in general
nonconvex (possibly NP hard \cite{Luo2008}), the focus of this paper
is to design \emph{distributed }solution methods for computing stationary
solutions (possibly local minima) of (\ref{eq:social-formulation}).
Our major goal is to devise \emph{parallel (nonlinear) best-response
}schemes that converge even when the expected value in (\ref{eq:social-formulation})
cannot be computed accurately and only sample values are available.

\section{\label{sec:The-Stochastic-Distributed}A Novel Parallel Stochastic
Decomposition}

The social problem (\ref{eq:social-formulation}) faces two main issues:
i) the nonconvexity of the objective functions; and ii) the impossibility
to estimate accurately the expected value. To deal with these difficulties,
we propose a decomposition scheme that consists in solving a sequence
of \emph{parallel strongly convex} subproblems (one for each user),
where the objective function of user $i$ is obtained from $U(\mathbf{x})$
by replacing the expected value with a suitably chosen incremental
estimate of it and linearizing the nonconvex part. More formally,
at iteration $t+1$, user $i$ solves the following problem: given
$\mathbf{x}_{-i}^{t}$ and $\boldsymbol{\xi}^{t}$,\begin{subequations}\label{eq:approximation-problem}
\begin{equation}
\hat{\mathbf{x}}_{i}(\mathbf{x}^{t},\boldsymbol{\xi}^{t})\triangleq\underset{\mathbf{x}_{i}\in\mathcal{X}_{i}}{\arg\min}\;\hat{f}_{i}(\mathbf{x}_{i};\mathbf{x}^{t},\boldsymbol{\xi}^{t}),\label{eq:update-rule-x-hat}
\end{equation}
with the approximation function $\hat{f}_{i}(\mathbf{x}_{i};\mathbf{x}^{t},\boldsymbol{\xi}^{t})$
defined as
\begin{align}
\hat{f}_{i}(\mathbf{x}_{i};\mathbf{x}^{t},\boldsymbol{\xi}^{t}) & \triangleq\nonumber \\
\rho^{t} & {\displaystyle {\sum_{j\in\mathcal{C}_{i}^{t}}}}f_{j}(\mathbf{x}_{i},\mathbf{x}_{-i}^{t},\boldsymbol{\xi}^{t})+\rho^{t}\left\langle \mathbf{x}_{i}-\mathbf{x}_{i}^{t},\boldsymbol{\pi}_{i}(\mathbf{x}^{t},\boldsymbol{\xi}^{t})\right\rangle \nonumber \\
+ & (1-\rho^{t})\bigl\langle\mathbf{x}_{i}-\mathbf{x}_{i}^{t},\mathbf{f}_{i}^{t-1}\bigr\rangle+\tau_{i}\bigl\Vert\mathbf{x}_{i}-\mathbf{x}_{i}^{t}\bigr\Vert^{2};\label{eq:update-rule-approximation-function}
\end{align}
where the pricing vector $\boldsymbol{\pi}_{i}(\mathbf{x},\boldsymbol{\xi})$
is given by
\begin{equation}
\boldsymbol{\pi}_{i}\left(\mathbf{x}^{t},\boldsymbol{\xi}^{t}\right)\triangleq{\displaystyle {\sum_{j\in\overline{\mathcal{C}}_{i}^{t}}}}\nabla_{i}f_{j}\left(\mathbf{x}^{t},\boldsymbol{\xi}^{t}\right);\vspace{-.1cm}\label{eq:update-rule-pricing-vector}
\end{equation}
and $\mathbf{f}_{i}^{t}$ is an accumulation vector updated recursively
according to\vspace{-.2cm}
\begin{equation}
\mathbf{f}_{i}^{t}=(1-\rho^{t})\mathbf{f}_{i}^{t-1}+\rho^{t}\left(\boldsymbol{\pi}_{i}(\mathbf{x}^{t},\boldsymbol{\xi}^{t})+{\textstyle \sum_{j\in\mathcal{C}_{i}^{t}}}\nabla_{i}f_{j}(\mathbf{x}^{t},\boldsymbol{\xi}^{t})\right),\label{eq:update-rule-incremental-gradient}
\end{equation}
\end{subequations}with $\rho^{t}\in(0,1]$ being a sequence to be
properly chosen ($\rho^{0}=1$). The other symbols in (\ref{eq:approximation-problem})
are defined as follows:
\begin{itemize}
\item In (\ref{eq:update-rule-incremental-gradient}): $C_{i}^{t}$ is any
subset of $S_{i}^{t}\triangleq\{i\in\mathcal{I}_{f}:\; f_{i}(\bullet,\mathbf{x}_{-i}^{t},\boldsymbol{\xi}^{t})\textrm{ is convex on }\mathcal{X}_{i}\}$
that is the set of indices of functions that are convex in $\mathbf{x}_{i}$,
given $\mathbf{x}_{-i}^{t}$ and $\boldsymbol{\xi}^{t}$;
\item In (\ref{eq:update-rule-pricing-vector}): $\overline{\mathcal{C}}_{i}^{t}$
denotes the complement of $C_{i}^{t}$; it contains (at least) the
indices of functions that are nonconvex in $\mathbf{x}_{i}$, given
$\mathbf{x}_{-i}^{t}$ and $\boldsymbol{\xi}^{t}$;
\item In (\ref{eq:update-rule-pricing-vector})-(\ref{eq:update-rule-incremental-gradient}):$\nabla_{i}f_{j}(\mathbf{x},\boldsymbol{\xi})$
is the gradient of $f_{j}(\mathbf{x},\boldsymbol{\xi})$ w.r.t. $\mathbf{x}_{i}^{*}$
(the complex conjugate of $\mathbf{x}_{i}$). Since $f_{j}(\mathbf{x},\boldsymbol{\xi})$
is real-valued, $\nabla_{\mathbf{x}_{i}^{*}}f(\mathbf{x},\boldsymbol{\xi})=\nabla_{\mathbf{x}_{i}^{*}}f(\mathbf{x},\boldsymbol{\xi})^{*}=(\nabla_{\mathbf{x}_{i}}f(\mathbf{x},\boldsymbol{\xi}))^{*}$. 
\end{itemize}
Given $\hat{\mathbf{x}}_{i}(\mathbf{x},\boldsymbol{\xi})$, $\mathbf{x}$
is updated according to
\begin{equation}
\mathbf{x}_{i}^{t+1}=\mathbf{x}_{i}^{t}+\gamma^{t+1}(\hat{\mathbf{x}}_{i}(\mathbf{x}^{t},\boldsymbol{\xi}^{t})-\mathbf{x}_{i}^{t}),\; i=1,\ldots,K,\label{eq:update-rule-x}
\end{equation}
where $\gamma^{t}\in(0,1]$. It turns our that $\mathbf{x}^{t}$ is
a random vector depending on $\mathcal{F}^{t}$, the past history
of the algorithm up to iteration $t$:
\begin{equation}
\mathcal{F}^{t}\triangleq\bigl\{\mathbf{x}^{0},\ldots,\mathbf{x}^{t-1},\boldsymbol{\xi}^{0},\ldots,\boldsymbol{\xi}^{t-1},\gamma^{1},\ldots,\gamma^{t},\rho^{0},\ldots,\rho^{t}\bigr\};\label{eq:update-history}
\end{equation}
therefore $\hat{\mathbf{x}}(\mathbf{x}^{t},\boldsymbol{\xi}^{t})$
depends on $\mathcal{F}^{t}$ as well (we omit this dependence for
notational simplicity).

The subproblems (\ref{eq:update-rule-x-hat}) have an interesting
interpretation: each user solves a sample convex approximation of
the original nonconvex stochastic function. The first term in (\ref{eq:update-rule-approximation-function})
preserves the convex component (or a part of it, if $C_{i}^{t}\subset S_{i}^{t}$)
of the (instantaneous) social function. The second term in (\ref{eq:update-rule-approximation-function})$-$the
pricing vector $\boldsymbol{\pi}_{i}(\mathbf{x},\boldsymbol{\xi})$$-$comes
from the linearization of (at least) the nonconvex part. The vector
$\mathbf{f}_{i}^{t}$ in the third term represents the incremental
estimate of $\nabla_{\mathbf{x}^{\ast}}U(\mathbf{x}^{t})$ (whose
value is not available), as one can readily check by substituting
(\ref{eq:update-rule-pricing-vector}) into (\ref{eq:update-rule-incremental-gradient}):
\begin{equation}
\mathbf{f}_{i}^{t}=(1-\rho^{t})\mathbf{f}_{i}^{t-1}+\rho^{t}{\textstyle \sum_{j\in\mathcal{I}_{f}}\nabla_{i}f_{j}(\mathbf{x}^{t},\boldsymbol{\xi})}.\label{eq:update-rule-incremental-gradient-1}
\end{equation}
Roughly speaking, the goal of this third term is to estimate on-the-fly
the unknown $\nabla_{\mathbf{x}^{\ast}}U(\mathbf{x}^{t})$ by its
samples collected over the iterations; based on (\ref{eq:update-rule-incremental-gradient-1}),
such an estimate is expected to become more and more accurate as $t$
increases, provided that the sequence $\rho^{t}$ is properly chosen
(this statement is made rigorous shortly in Theorem \ref{thm:convergence}).
The last quadratic term in (\ref{eq:update-rule-approximation-function})
is the proximal regularization whose numerical benefits are well-understood
\cite{Bertsekas}.\textcolor{red}{{} }

Given (\ref{eq:approximation-problem}), we define the ``best-response''
mapping as
\begin{equation}
\mathcal{X}\ni\mathbf{y}\mapsto\hat{\mathbf{x}}(\mathbf{y},\boldsymbol{\xi})\triangleq\left(\hat{\mathbf{x}}_{i}(\mathbf{y},\boldsymbol{\xi})\right)_{i=1}^{I}.\label{eq:update-mapping}
\end{equation}
Note that $\widehat{\mathbf{x}}(\bullet,\boldsymbol{\xi})$ is well-defined
for any given $\boldsymbol{\xi}$ because the objective function in
(\ref{eq:approximation-problem}) is strongly convex with constant
$\tau_{\min}$:
\begin{equation}
\begin{split}\tau_{\min}\triangleq & \min_{i=1,\ldots,I}\left\{ \tau_{i}\right\} .\end{split}
\label{eq:tau_min}
\end{equation}

\begin{algorithm}[H]
\textbf{Data}: $\boldsymbol{\tau}\triangleq(\tau_{i})_{i=1}^{I}\geq\mathbf{0}$,
$\left\{ \gamma^{t}\right\} $, $\left\{ \rho^{t}\right\} $, $\mathbf{x}^{0}\in\mathcal{X}$;
set $t=0$.

$(\texttt{S.1})$: If $\mathbf{x}^{t}$ satisfies a suitable termination
criterion: STOP.

$(\texttt{S.2})$: For all $i=1,\ldots,I$, compute $\widehat{\mathbf{x}}_{i}(\mathbf{x}^{t},\boldsymbol{\xi}^{t})$
{[}cf. (\ref{eq:approximation-problem}){]}.

$(\texttt{S.3})$: For all $i=1,\ldots,I$, update $\mathbf{x}^{t+1}$
according to 
\[
\mathbf{x}_{i}^{t+1}=(1-\gamma^{t+1})\mathbf{x}_{i}^{t}+\gamma^{t+1}\,\widehat{\mathbf{x}}_{i}(\mathbf{x}^{t},\boldsymbol{\xi}^{t}).
\]

$(\texttt{S.4})$: For all $i=1,\ldots,I$, update $\mathbf{f}_{i}^{t}$
according to (\ref{eq:update-rule-incremental-gradient}).

$(\texttt{S.5})$: $t\leftarrow t+1$, and go to $(\texttt{S.1})$.

\protect\caption{\hspace{-2.5pt}\textbf{:} \label{alg}Stochastic parallel decomposition
algorithm}
\end{algorithm}
 Our decomposition scheme is formally described in Algorithm \ref{alg},
and its convergence properties are stated in Theorem \ref{thm:convergence},
under the following standard boundedness assumptions on the instantaneous
gradient errors \cite{Zhang2008b,Ram2009}:\vspace{0.5em}

\noindent \textbf{Assumption }(c): The instantaneous gradient is
unbiased with bounded variance in the following sense:
\[
\mathbb{E}\bigl[\nabla U(\mathbf{x}^{t})-{\textstyle \sum_{j\in\mathcal{I}_{f}}}\nabla f_{j}(\mathbf{x}^{t},\boldsymbol{\xi}^{t})\bigr|\mathcal{F}^{t}\bigr]=\mathbf{0},\; t=0,1,\ldots
\]
and
\[
\mathbb{E}\bigl[\bigl\Vert\nabla U(\mathbf{x}^{t})-{\textstyle \sum_{j\in\mathcal{I}_{f}}}\nabla f_{j}(\mathbf{x}^{t},\boldsymbol{\xi}^{t})\bigr\Vert^{2}\bigl|\mathcal{F}^{t}\bigr]<\infty,\; t=0,1,\ldots
\]
This assumption is readily satisfied if $\boldsymbol{\xi}$ is a bounded
i.i.d. random variable.
\begin{thm}
\label{thm:convergence}Given problem (\ref{eq:social-formulation})
under Assumptions (a)-(c), suppose that $\tau_{\min}>0$ and the stepsizes
$\left\{ \gamma^{t}\right\} $ and $\left\{ \rho^{t}\right\} $ are
chosen so that \begin{subequations}\label{eq:stepsize}
\begin{align}
\mathbf{i)} & \;\gamma^{t}\rightarrow0,\;{\textstyle \sum_{t}}\,\gamma^{t}=\infty,\;{\textstyle \sum}\,(\gamma^{t})^{2}<\infty,\label{eq:stepsize-1}\\
\mathbf{ii)} & \;\rho^{t}\rightarrow0,\;{\textstyle \sum_{t}}\,\rho^{t}=\infty,\;{\textstyle \sum}\,(\rho^{t})^{2}<\infty,\label{eq:stepsize-2}\\
\mathbf{iii)} & \;\lim_{t\rightarrow\infty}\gamma^{t}/\rho^{t}=0,\label{eq:stepsize-3}\\
\mathbf{iv)} & \;\underset{t\rightarrow\infty}{\lim\sup}\;\rho^{t}\Bigl({\textstyle \sum}_{j\in\mathcal{I}_{f}}L_{\nabla f_{j}(\boldsymbol{\xi}^{t})}\Bigr)=0,\;\mathrm{a.s}.\label{eq:stepsize-4}
\end{align}
\end{subequations}Then, every limit point of the sequence $\left\{ \mathbf{x}^{t}\right\} $
generated by Algorithm \ref{alg} (at least one of such point exists)
is a stationary point of (\ref{eq:social-formulation}) almost surely.\end{thm}
\begin{IEEEproof}
See Appendix \ref{sec:Proof-of-convergence}.
\end{IEEEproof}
\noindent \textbf{On Assumption (c): }The boundedness condition is
in terms of the conditional expectation of the (random) gradient error.
Compared with \cite{Razaviyayn_2013}, Assumption (c) is weaker because
it is required in \cite{Razaviyayn_2013} that every realization of
the (random) gradient error must be bounded.

\noindent \textbf{On Condition (\ref{eq:stepsize-4}): }The condition
has the following interpretation: all increasing subsequences of $\sum_{j\in\mathcal{I}_{f}}L_{\nabla f_{j}(\boldsymbol{\xi}^{t})}$
must grow slower than $1/\rho^{t}$. We will discuss later in Sec.
\ref{sec:applications} how this assumption is satisfied by specific
applications. Note that if $\sum_{j\in\mathcal{I}_{f}}L_{\nabla f_{j}(\boldsymbol{\xi})}$
is uniformly bounded for any $\boldsymbol{\xi}$ (which is indeed
the case if $\boldsymbol{\xi}$ is a bounded random vector), then
(\ref{eq:stepsize-4}) is trivially satisfied.

\noindent \textbf{On Algorithm \ref{alg}: }To our best knowledge,
Algorithm \ref{alg} is the first \emph{parallel} \emph{best-response}
(e.g., nongradient-like) scheme for nonconvex stochastic social problems:
all the users update \emph{in parallel }their strategies (possibly
with a memory) solving a sequence of \emph{decoupled} (strongly) convex
subproblems (\ref{eq:approximation-problem}). It is expected to perform
better than classical stochastic gradient-based schemes at no the
cost of extra signaling, because the convexity of the objective function,
if any, is better exploited. Our experiments on specific applications
confirm this intuition; see Sec. \ref{sec:applications}. Moreover,
it is guaranteed to converge under the weakest assumptions available
in literature while offering some flexibility in the choice of the
free parameters {[}cf. Theorem \ref{thm:convergence}{]}.

\noindent \textbf{Diminishing stepsize rules: }In order to have convergence,
a diminishing stepsize rule satisfying (\ref{eq:stepsize}) is necessary.
An instance of (\ref{eq:stepsize}) is, e.g., the following: 
\begin{equation}
\gamma^{t}=\frac{1}{t^{\alpha}},\;\rho^{t}=\frac{1}{t^{\beta}},\;0.5<\beta<\alpha\leq1.\label{eq:ex_step_size}
\end{equation}
Roughly speaking, (\ref{eq:stepsize}) says that the stepsizes $\gamma^{t}$
and $\rho^{t}$, while diminishing (with $\gamma^{t}$ decreasing
faster than $\rho^{t}$), need not go to zero too fast. This kind
of stepsize rules are of the same spirit of those used to guarantee
convergence of gradient methods with error; see \cite{Srivastava2011}
for more details.

\noindent \textbf{Implementation issues: }In order to compute the
best-response, each user needs to know $\sum_{j\in\mathcal{C}_{i}^{t}}\, f_{j}(\mathbf{x}_{i},\mathbf{x}_{-i}^{t},\boldsymbol{\xi}^{t})$
and the pricing vector $\boldsymbol{\pi}_{i}(\mathbf{x}^{t},\boldsymbol{\xi}^{t})$.
The signaling required to acquire this information is of course problem-dependent.
If the problem under consideration does not have any specific structure,
the most natural message-passing strategy is to communicate directly
$\mathbf{x}_{-i}^{t}$ and $\boldsymbol{\pi}_{i}(\mathbf{x}^{t},\boldsymbol{\xi}^{t})$.
However, in many specific applications much less signaling may be
needed; see Sec. \ref{sec:applications} for some examples. Note that
the signaling is of the same spirit of that of pricing-based algorithms
proposed in the literature for the maximization of deterministic sum-utility
functions \cite{Kim2011,Scutarib}; no extra communication is required
to update $\mathbf{f}_{i}^{t}$: once the new pricing vector $\boldsymbol{\pi}_{i}(\mathbf{x}^{t},\boldsymbol{\xi}^{t})$
is available, the recursive update (\ref{eq:update-rule-incremental-gradient})
for the ``incremental'' gradient is based on a local accumulation
register keeping track of the last iterate $\mathbf{f}_{i}^{t-1}$.
Note also that, thanks to the simultaneous nature of the proposed
scheme, the overall communication overhead is expected to be less
than that required to implement \emph{sequential }schemes, such as
\cite{Kim2011}. \vspace{-0.2cm}

\subsection{Some special cases}

We customize next the proposed general algorithmic framework to specific
classes of problems (\ref{eq:social-formulation}) arising naturally
in many applications.

\subsubsection{Stochastic proximal conditional gradient methods}

Quite interestingly, the proposed decomposition technique resembles
classical stochastic conditional gradient schemes \cite{polyak1987introduction}
when one chooses in (\ref{eq:update-rule-approximation-function})
$\mathcal{C}_{i}^{t}=\emptyset$, for all $i$ and $t$, resulting
in the following approximation function:
\begin{align}
\hat{f}_{i}(\mathbf{x}_{i};\mathbf{x}^{t},\boldsymbol{\xi}^{t}) & =\rho^{t}\bigl\langle\mathbf{x}_{i}-\mathbf{x}_{i}^{t},{\textstyle \sum_{j\in\mathcal{I}_{f}}}\nabla_{i}f_{j}\left(\mathbf{x}^{t},\boldsymbol{\xi}^{t}\right)\bigr\rangle\nonumber \\
+(1-\rho^{t}) & \bigl\langle\mathbf{x}_{i}-\mathbf{x}_{i}^{t},\mathbf{f}_{i}^{t-1}\bigr\rangle+\tau_{i}\left\Vert \mathbf{x}_{i}-\mathbf{x}_{i}^{t}\right\Vert ^{2},\label{eq:conditional-gradient}
\end{align}
with $\mathbf{f}_{i}^{t}$ updated according to (\ref{eq:update-rule-incremental-gradient-1}).
Note that, however, traditional stochastic conditional gradient methods
\cite{Ermol'ev1977} do not have the proximal regularization term
in (\ref{eq:conditional-gradient}), which instead brings in well-understood
numerical benefits. Moreover, it is worth mentioning that, for some
of the applications introduced in Sec. \ref{sec:Problem-Formulation},
it is just the presence of the proximal term that allows one to compute
the best-response $\hat{\mathbf{x}}_{i}(\mathbf{x}^{t},\boldsymbol{\xi}^{t})$
resulting from the minimization of (\ref{eq:conditional-gradient})
in closed-form; see Sec. \ref{sub:application:MIMO-IC}.

It turns out that convergence conditions of Algorithm \ref{alg} contain
as special cases those of classical stochastic conditional gradient
methods. But the proposed algorithmic framework is much more general
and, among all,  is able to better exploit the structure of the sum-utility
function (if any) than just linearizing everything; it is thus expected
to be faster than classical stochastic conditional gradient methods,
a fact that is confirmed by our experiments; see Sec. \ref{sec:applications}.

\subsubsection{Stochastic best-response algorithm for single (convex) functions}

Suppose that the social function is a single function $U(\mathbf{x})=\mathbb{E}\left[f(\mathbf{x}_{1},\ldots,\mathbf{x}_{I},\boldsymbol{\xi})\right]$
on $\mathcal{X}=\prod_{i}\mathcal{X}_{i}$, and $f(\mathbf{x}_{1},\ldots,\mathbf{x}_{I},\boldsymbol{\xi})$
is uniformly convex in each $\mathbf{x}_{i}$ (but not necessarily
jointly). Of course, this optimization problem can be interpreted
as a special case of the framework (\ref{eq:social-formulation}),
with $I_{f}=1$ and $\mathcal{I}_{f}=\{1\}$ and $S_{i}^{t}=\{1\}$.
Since $f(\bullet,\boldsymbol{\xi})$ is already convex separately
in the variables $\mathbf{x}_{i}$'s, a natural choice for the approximants
$\hat{f}_{i}$ is setting $\mathcal{C}_{i}^{t}=S_{i}^{t}=\{1\}$ for
all $t$, resulting in the following:
\begin{align}
\hat{f}_{i}(\mathbf{x}_{i};\mathbf{x}^{t},\boldsymbol{\xi}^{t}) & =\;\rho^{t}f\left(\mathbf{x}_{i},\mathbf{x}_{-i}^{t},\boldsymbol{\xi}^{t}\right)\nonumber \\
+(1-\rho^{t}) & \bigl\langle\mathbf{x}_{i}-\mathbf{x}_{i}^{t},\mathbf{f}_{i}^{t-1}\bigr\rangle+\tau_{i}\left\Vert \mathbf{x}_{i}-\mathbf{x}_{i}^{t}\right\Vert ^{2},\label{eq:convex-jacobi}
\end{align}
where $\mathbf{f}_{i}^{t}$ is updated according to $\mathbf{f}_{i}^{t}=\left(1-\rho^{t}\right)\mathbf{f}_{i}^{t-1}+\rho^{t}\nabla_{i}f\left(\mathbf{x}^{t},\boldsymbol{\xi}^{t}\right)$.
Convergence conditions are still given by Theorem \ref{thm:convergence}.
It is worth mentioning that the same choice comes out naturally when
$f(\mathbf{x}_{1},\ldots,\mathbf{x}_{I},\boldsymbol{\xi})$ is uniformly
\emph{jointly }convex; in such a case the proposed algorithm converges
(in the sense of Theorem \ref{thm:convergence}) to the \emph{global
optimum} of $U(\mathbf{x})$. An interesting application of this algorithm
is the maximization of the ergodic sum-rate over MIMO MACs in (\ref{eq:MIMO-MAC-SR}),
resulting in the first \emph{convergent} \emph{simultaneous }stochastic
MIMO Iterative Waterfilling algorithm in the literature; see Sec.
\ref{sub:MIMO-MAC}.

\subsubsection{Stochastic pricing algorithms}

Suppose that $I=I_{f}$ and each $S_{i}^{t}=\{i\}$ (implying that
$f_{i}(\bullet,\mathbf{x}_{-i},\boldsymbol{\xi})$ is uniformly convex
on $\mathcal{X}_{i}$). By taking each $\mathcal{C}_{i}^{t}=\{i\}$
for all $t$, the approximation function in (\ref{eq:update-rule-approximation-function})
reduces to
\begin{equation}
\hspace{-0.3cm}\begin{array}{l}
\hat{f}_{i}(\mathbf{x}_{i};\mathbf{x}^{t},\boldsymbol{\xi}^{t})\triangleq\;\rho^{t}f_{i}(\mathbf{x}_{i},\mathbf{x}_{-i}^{t},\boldsymbol{\xi}^{t})+\rho^{t}\bigl\langle\mathbf{x}_{i}-\mathbf{x}_{i}^{t},\boldsymbol{\pi}_{i}(\mathbf{x}^{t},\boldsymbol{\xi}^{t})\bigr\rangle\smallskip\\
\qquad\qquad\qquad\quad+(1-\rho^{t})\bigl\langle\mathbf{x}_{i}-\mathbf{x}_{i}^{t},\mathbf{f}_{i}^{t-1}\bigr\rangle+\tau_{i}\left\Vert \mathbf{x}_{i}-\mathbf{x}_{i}^{t}\right\Vert ^{2},
\end{array}\label{eq:partial-linearization}
\end{equation}
where $\boldsymbol{\pi}_{i}(\mathbf{x},\boldsymbol{\xi})=\sum_{j\neq i}\nabla_{i}f_{j}(\mathbf{x},\boldsymbol{\xi})$
and $\mathbf{f}_{i}^{t}=\left(1-\rho^{t}\right)\mathbf{f}_{i}^{t-1}+\rho^{t}(\boldsymbol{\pi}_{i}(\mathbf{x}^{t},\boldsymbol{\xi}^{t})+\nabla_{i}f_{i}(\mathbf{x}_{i},\mathbf{x}_{-i}^{t},\boldsymbol{\xi}^{t}))$.
This is the generalization of the deterministic pricing algorithms
\cite{Kim2011,Scutarib} to stochastic optimization problems. Examples
of this class of problems are the ergodic sum-rate maximization problem
over SISO and MIMO IC in (\ref{eq:SISO-IC-SR})-(\ref{eq:MIMO-IC-SR});
see Sec. \ref{sub:SISO-IC} and Sec. \ref{sub:application:MIMO-IC}.

\subsubsection{Stochastic DC programming}

A stochastic DC programming problem is formulated as
\begin{equation}
\begin{array}{cl}
\underset{\mathbf{x}}{\textrm{minimize}} & \mathbb{E}_{\boldsymbol{\xi}}\left[\sum_{j\in\mathcal{I}_{f}}\left(f_{j}(\mathbf{x},\boldsymbol{\xi})-g_{j}(\mathbf{x},\boldsymbol{\xi})\right)\right]\smallskip\\
\textrm{subject to} & \mathbf{x}_{i}\in\mathcal{X}_{i},\quad i=1,\ldots,I,
\end{array}\label{eq:DC-Programming}
\end{equation}
where both $f_{j}(\mathbf{\bullet},\boldsymbol{\xi})$ and $g_{j}(\mathbf{\bullet},\boldsymbol{\xi})$
are uniformly convex functions on $\mathcal{X}$. A natural choice
of the approximation functions $\hat{f}_{i}$ for (\ref{eq:DC-Programming})
is linearizing the concave part of the sample social function, resulting
in the following:
\[
\hspace{-0.3cm}\begin{split}\hat{f}_{i}(\mathbf{x}_{i};\mathbf{x}^{t},\boldsymbol{\xi}^{t})= & \rho^{t}{\textstyle \sum_{j\in\mathcal{I}_{f}}}f_{j}(\mathbf{x}_{i},\mathbf{x}_{-i}^{t},\boldsymbol{\xi}^{t})+\rho^{t}\bigl\langle\mathbf{x}_{i}-\mathbf{x}_{i}^{t},\boldsymbol{\pi}_{i}(\mathbf{x}^{t},\boldsymbol{\xi}^{t})\bigr\rangle\\
 & +(1-\rho^{t})\bigl\langle\mathbf{x}_{i}-\mathbf{x}_{i}^{t},\mathbf{f}_{i}^{t-1}\bigr\rangle+\tau_{i}\,\bigl\Vert\mathbf{x}_{i}-\mathbf{x}_{i}^{t}\bigr\Vert^{2},
\end{split}
\]
where $\boldsymbol{\pi}_{i}(\mathbf{x},\boldsymbol{\xi})\triangleq-\sum_{j\in\mathcal{I}_{f}}\nabla_{i}g_{j}(\mathbf{x},\boldsymbol{\xi})$
and $\mathbf{f}_{i}^{t}=\left(1-\rho^{t}\right)\mathbf{f}_{i}^{t-1}+\rho^{t}\left(\boldsymbol{\pi}_{i}(\mathbf{x}^{t},\boldsymbol{\xi}^{t})+\sum_{j\in\mathcal{I}_{f}}\nabla_{i}f_{j}(\mathbf{x}_{i},\mathbf{x}_{-i}^{t},\boldsymbol{\xi}^{t})\right)$.\vspace{-0.2cm}

\section{\label{sec:applications}Applications}

We customize now the proposed algorithmic framework to some of the
applications introduced in Sec. \ref{sec:Problem-Formulation}, and
compare the resulting algorithms with state-of-the-art schemes proposed
for the specific problems under considerations as well as more classical
stochastic gradient algorithms. Numerical results provide a solid
evidence of the superiority of our approach. \vspace{-0.3cm}

\subsection{\label{sub:SISO-IC}Sum-rate maximization over frequency-selective
ICs}

Consider the sum-rate maximization problem over frequency-selective
ICs, as introduced in (\ref{eq:SISO-IC-SR}). Since the instantaneous
rate of each user $i$,
\[
r_{i}(\mathbf{p}_{i},\mathbf{p}_{-i},\mathbf{h})=\sum_{n=1}^{N}\log\left(1+\frac{|h_{ii,n}|^{2}\, p_{i,n}}{\sigma_{i,n}^{2}+\sum_{j\neq i}|h_{ij,n}|^{2}\, p_{j,n}}\right),
\]
is uniformly strongly concave in $\mathbf{p}_{i}\in\mathcal{P}_{i}$,
a natural choice for the approximation function $\hat{{f}}_{i}$ is
the one in (\ref{eq:partial-linearization}) wherein $r_{i}(\mathbf{p}_{i},\mathbf{p}_{-i},\mathbf{h}^{t})$
is not touched while $\sum_{j\neq i}r_{j}(\mathbf{p}_{j},\mathbf{p}_{-j},\mathbf{h}^{t})$
is linearized. This leads to the following best-response functions
\begin{subequations}
\begin{align}
\hat{\mathbf{p}}_{i}(\mathbf{p}^{t},\mathbf{h}^{t}) & =\underset{\mathbf{p}_{i}\in\mathcal{P}_{i}}{\arg\max}\Bigl\{\rho^{t}\cdot r_{i}(\mathbf{p}_{i},\mathbf{p}_{-i}^{t},\mathbf{h}^{t})+\rho^{t}\left<\mathbf{p}_{i},\boldsymbol{\pi}_{i}^{t}\right>\nonumber \\
+ & (1-\rho^{t})\left<\mathbf{p}_{i},\mathbf{f}_{i}^{t-1}\right>-\frac{\tau_{i}}{2}\left\Vert \mathbf{p}_{i}-\mathbf{p}_{i}^{t}\right\Vert ^{2}\Bigr\},\label{eq:SISO-p-hat}
\end{align}
where $\boldsymbol{\pi}_{i}^{t}$$=$$\boldsymbol{\pi}_{i}(\mathbf{p}^{t},\mathbf{h}^{t})\triangleq(\pi_{i,n}(\mathbf{p}^{t},\mathbf{h}^{t}))_{n=1}^{N}$
with
\[
\begin{split}\pi_{i,n}(\mathbf{p}^{t},\mathbf{h}^{t}) & =\sum_{j\neq i}\nabla_{p_{i,n}}r_{j}(\mathbf{p}^{t},\mathbf{h}^{t})\\
 & =-\sum_{j\neq i}|h_{ji,n}^{t}|^{2}\frac{\texttt{SINR}_{j,n}^{t}}{(1+\texttt{SINR}_{j,n}^{t})\cdot\texttt{MUI}_{j,n}^{t}},\\
\texttt{MUI}_{j,n}^{t} & \triangleq\sigma_{j,n}^{2}+\sum_{i\neq j}|h_{ji,n}^{t}|^{2}p_{i,n}^{t},\\
\texttt{SINR}_{j,n}^{t} & =|h_{jj,n}^{t}|^{2}p_{j,n}^{t}/\texttt{MUI}_{j,n}^{t}.
\end{split}
\]
\end{subequations} Then the variable $\mathbf{f}_{i}^{t}$ is updated
according to $\mathbf{f}_{i}^{t}=(1-\rho^{t})\mathbf{f}_{i}^{t-1}+\rho^{t}(\boldsymbol{\pi}_{i}^{t}+\nabla_{\mathbf{p}_{i}}r_{i}(\mathbf{p}^{t},\mathbf{h}^{t})).$
Note that the optimization problem in (\ref{eq:SISO-p-hat}) has a
closed-form expression \cite{Scutarib}:
\begin{align}
\hat{p}_{i,n}(\mathbf{p}^{t},\mathbf{h}^{t})=\; & \texttt{WF}\bigl(\rho^{t},\texttt{SINR}_{i,n}^{t}/p_{i,n}^{t},\tau_{i},\nonumber \\
 & \rho^{t}\pi_{i,n}^{t}+(1-\rho^{t})f_{i,n}^{t-1}+\tau_{i}^{t}p_{i,n}^{t}-\mu^{\star}\bigr),\label{eq:best-response-power}
\end{align}
where\vspace{-0.2cm}
\[
\texttt{WF}(a,b,c,d)=\frac{1}{2}\left[\frac{d}{c}-\frac{1}{b}+\sqrt{\left(\frac{d}{c}+\frac{1}{b}\right)^{2}+\frac{4a}{c}}\right]^{+},
\]
and $\mu^{\star}$ is the Lagrange multiplier such that $0\leq\mu^{\star}\perp\sum_{n=1}^{N}\hat{p}_{i,n}(\mathbf{p}^{t},\mathbf{h}^{t})-P_{i}\leq0$,
and it can be found by standard bisection method.

The overall stochastic pricing-based algorithm is then given by Algorithm
1 with best-response mapping defined in (\ref{eq:best-response-power});
convergence is guaranteed under conditions i)-iv) in Theorem \ref{thm:convergence}.
Note that the theorem is trivially satisfied using stepsizes rules
as required in i)-iii) {[}e.g., (\ref{eq:ex_step_size}){]}; the only
condition that needs some comment is condition iv). If $\underset{t\rightarrow\infty}{\lim\sup}\;\rho^{t}\Bigl({\textstyle \sum}_{j\in\mathcal{I}_{f}}L_{\nabla f_{j}(\boldsymbol{\xi}^{t})}\Bigr)>0,$
we can assume without loss of generality (w.l.o.g.) that the sequence
of the Lipschitz constant $\bigl\{{\textstyle \sum}_{j\in\mathcal{I}_{f}}L_{\nabla f_{j}(\boldsymbol{\xi}^{t})}\bigr\}$
is increasing monotonically at a rate no slower than $1/\rho^{t}$
(we can always limit the discussion to such a subsequence). For any
$\bar{h}>0$, define $p(\bar{h})\triangleq\text{{Prob}(}|h_{ij,n}|\geq\bar{h})$
and assume w.l.o.g. that $0\leq p(\bar{h})<1$. Note that the Lipschitz
constant $L_{\nabla f_{j}(\boldsymbol{\xi})}$ is upper bounded by
the maximum eigenvalue of the augmented Hessian of $f_{j}(\mathbf{x},\boldsymbol{\xi})$
\cite{bertsekas1999nonlinear}, and the maximum eigenvalue increasing
monotonically means that the channel coefficient is becoming larger
and larger (this can be verified by explicitly calculating the augmented
Hessian of $f_{j}(\mathbf{x},\boldsymbol{\xi})$; detailed steps are
omitted due to page limit). Since $\textrm{Prob}(|h_{ij,n}^{t+1}|\geq|h_{ij,n}^{t}|\textrm{ for all }t\geq t_{0})\leq\textrm{Prob}(|h_{ij,n}^{t+1}|\geq\bar{h}\textrm{ for all }t\geq t_{0})=p(\bar{h})^{t-t_{0}+1}\rightarrow0$,
we can infer that the magnitude of the channel coefficient increasing
monotonically is an event of probability 0. Therefore, condition (\ref{eq:stepsize-4})
is satisfied.
\begin{figure*}[tbh]
\center\subfigure[ergodic sum-rate versus iterations]{\includegraphics[scale=0.7]{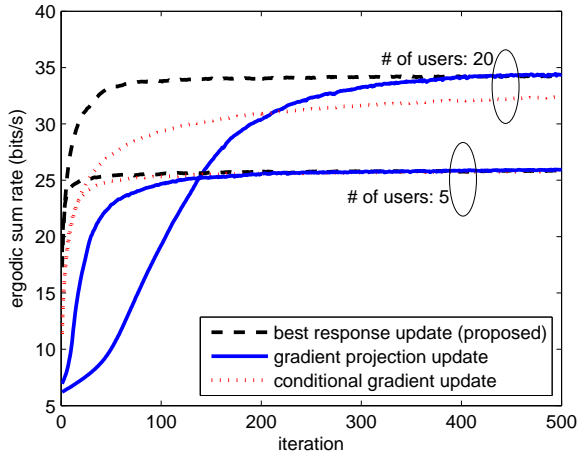}}\hspace{0.5cm}\subfigure[achievable sum-rate versus iterations]{\includegraphics[scale=0.7]{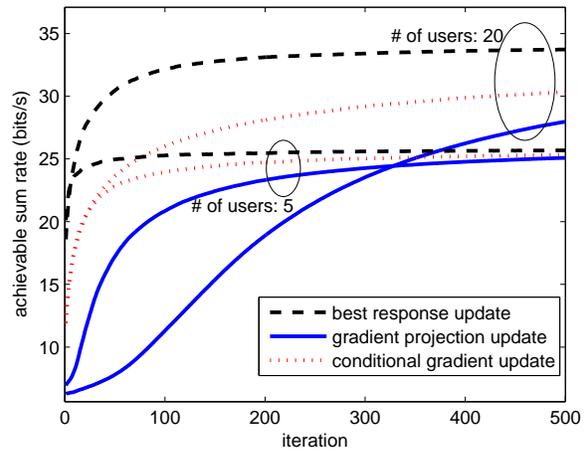}}\protect\caption{\label{fig:SISO-IC-SR}Sum-rate versus iteration in frequency-selective
ICs.}
\vspace{-0.5cm}
\end{figure*}
 \\\noindent\textbf{Numerical results}\emph{. }We simulated a SISO
frequency selective IC under the following setting: the number of
users is either five or twenty; equal power budget $P_{i}=P$ and
white Gaussian noise variance $\sigma_{i}^{2}=\sigma^{2}$ are assumed
for all users; the SNR of each user $\texttt{snr}=P/\sigma^{2}$ is
set to $10$dB; the instantaneous parallel subchannels $\mathbf{h}^{t}\triangleq(h_{ij,n}^{t})_{i,j,n}$
are generated according to $\mathbf{h}^{t}=\mathbf{h}+\triangle\mathbf{h}^{t}$,
where $\mathbf{h}$ (generated by MATLAB command $\texttt{randn}$)
is fixed while $\triangle\mathbf{h}^{t}$ is generated at each $t$
using $\delta\cdot\texttt{randn}$, with $\delta=0.2$ being the noise
level. We simulated the following algorithms: i) the proposed stochastic
best-response pricing algorithm (with $\tau_{i}=10^{-8}$ for all
$i$, $\gamma^{1}=\rho^{0}=\rho^{1}=1$, $\rho^{t}=2/(t+2)^{0.6}$,
and $\gamma^{t}=2/(t+2)^{0.61}$ for $t\geq2$); ii) the stochastic
conditional gradient method \cite{Ermol'ev1977} (with $\gamma^{1}=\rho^{0}=\rho^{1}=1$,
$\rho^{t}=1/(t+2)^{0.9}$, and $\gamma^{t}=1/(t+2)^{0.91}$ for $t\geq2$);
iii) and the stochastic gradient projection method \cite{DiLorenzo2013}
(with $\gamma^{1}=1$ and $\gamma^{t}=\gamma^{t-1}(1-10^{-3}\gamma^{t-1})$
for $t\geq2$). Note that the stepsizes are tuned such that all algorithms
can achieve their best empirical convergence speed. We plot two merit
functions, namely: i) the ergodic sum-rate, defined as $\mathbb{E}_{\mathbf{h}}[\sum_{n=1}^{N}\sum_{i=1}^{I}r_{i}(\mathbf{p}^{t},\mathbf{h})]$
(with the expected value estimated by the sample mean of 1000 independent
realizations); and ii) the ``achievable'' sum-rate, defined as$\frac{1}{t}\sum_{m=1}^{t}\sum_{n=1}^{N}\sum_{i=1}^{I}r_{i}(\mathbf{p}^{m},\mathbf{h}^{m}),$
which represents the sum-rate that is actually achieved in practice
(it is the time average of the instantaneous (random) sum-rate). \\ \indent
In Figure \ref{fig:SISO-IC-SR}, we plot the two above merit functions
versus the iteration index $t$ achieved using the different algorithms.
Our experiment show that for ``small'' systems (e.g., five active
users), all algorithms perform quite well (for both the merit functions),
with a gain in convergence speed for the proposed scheme. However,
when the number of users increases increased (e.g., from 5 to 20),
all other (gradient-like) algorithms suffer from very slow convergence.
Quite interestingly, the proposed scheme seems also quite scalable:
the convergence speed is not notably affected by the number of users,
which makes it applicable to more realistic scenarios. The faster
convergence of proposed stochastic best-response pricing algorithm
comes from a better exploitation of partial convexity in the problem
than what more classical gradient algorithms do, which validates the
main idea of this paper.\vspace{-0.3cm}

\subsection{\label{sub:application:MIMO-IC}Sum-rate maximization over MIMO ICs}

Now we customize Algorithm \ref{alg} to solve the sum-rate maximization
problem over MIMO ICs (\ref{eq:MIMO-IC-SR}). Defining
\[
r_{i}(\mathbf{Q}_{i},\mathbf{Q}_{-i},\mathbf{H})\triangleq\log\det\left(\mathbf{I}+\mathbf{H}_{ii}\mathbf{Q}_{i}\mathbf{H}_{ii}^{H}\mathbf{R}_{i}(\mathbf{Q}_{-i},\mathbf{H})^{-1}\right)
\]
and following a similar approach as in the SISO case, the best-response
of each user $i$ becomes {[}cf. (\ref{eq:partial-linearization}){]}:\vspace{-0.1cm}\begin{subequations}\label{eq:IC-update}
\begin{align}
\hat{\mathbf{Q}}_{i}(\mathbf{Q}^{t},\mathbf{H}^{t}) & =\underset{\mathbf{Q}_{i}\in\mathcal{Q}_{i}}{\arg\max}\Bigl\{\rho^{t}r_{i}(\mathbf{Q}_{i},\mathbf{Q}_{-i}^{t},\mathbf{H}^{t})+\rho^{t}\bigl\langle\mathbf{Q}_{i}-\mathbf{Q}_{i}^{t},\boldsymbol{\Pi}_{i}^{t}\bigr\rangle\nonumber \\
+(1-\rho^{t}) & \bigl\langle\mathbf{Q}_{i}-\mathbf{Q}_{i}^{t},\mathbf{F}_{i}^{t-1}\bigr\rangle-\tau_{i}\bigl\Vert\mathbf{Q}_{i}-\mathbf{Q}_{i}^{t}\bigr\Vert^{2}\Bigr\},\label{eq:IC-Q-hat}
\end{align}
where $\bigl\langle\mathbf{A},\mathbf{B}\bigr\rangle\triangleq\textrm{tr}(\mathbf{A}^{H}\mathbf{B})$;
$\boldsymbol{\Pi}_{i}\left(\mathbf{Q},\mathbf{H}\right)$ is given
by
\begin{equation}
\boldsymbol{\Pi}_{i}\left(\mathbf{Q},\mathbf{H}\right)=\sum_{j\neq i}\nabla_{\mathbf{Q}_{i}^{*}}r_{j}(\mathbf{Q},\mathbf{H})=\sum_{j\neq i}\mathbf{H}_{ji}^{H}\,\widetilde{\mathbf{R}}_{j}\left(\mathbf{Q}_{-j},\mathbf{H}\right)\mathbf{H}_{ji},\label{eq:IC-pricing}
\end{equation}
where $r_{j}(\mathbf{Q},\mathbf{H})=\log\det(\mathbf{I}+\mathbf{H}_{jj}\mathbf{Q}_{j}\mathbf{H}_{jj}^{H}\mathbf{R}_{j}(\mathbf{Q}_{-i},\mathbf{H})^{-1})$
and $\widetilde{\mathbf{R}}_{j}\left(\mathbf{Q}_{-j},\mathbf{H}\right)\triangleq\left(\mathbf{R}_{j}\left(\mathbf{Q}_{-j},\mathbf{H}\right)+\mathbf{H}_{jj}\mathbf{Q}_{j}\mathbf{H}_{jj}^{H}\right)^{-1}-\mathbf{R}_{j}\left(\mathbf{Q}_{-j},\mathbf{H}\right)^{-1}$.
Then $\mathbf{F}_{i}^{t}$ is updated by (\ref{eq:update-rule-incremental-gradient}),
which becomes
\begin{align}
\mathbf{F}_{i}^{t} & =(1-\rho^{t})\,\mathbf{F}_{i}^{t-1}+\rho^{t}{\textstyle \sum_{j=1}^{I}}\nabla_{\mathbf{Q}_{i}^{*}}r_{j}(\mathbf{Q}^{t},\mathbf{H}^{t})\nonumber \\
 & =(1-\rho^{t})\,\mathbf{F}_{i}^{t-1}+\rho^{t}\boldsymbol{\Pi}_{i}(\mathbf{Q}^{t},\mathbf{H}^{t})\nonumber \\
 & \quad\,+\rho^{t}(\mathbf{H}_{ii}^{t})^{H}\left(\mathbf{R}_{i}^{t}+\mathbf{H}_{ii}^{t}\mathbf{Q}_{i}^{t}(\mathbf{H}_{ii}^{t})^{H}\right)^{-1}\mathbf{H}_{ii}^{t}.\label{eq:IC-F}
\end{align}
 \end{subequations}We can then apply Algorithm \ref{alg} based on
the best-response $\hat{\mathbf{Q}}(\mathbf{Q}^{t},\mathbf{H}^{t})=(\hat{\mathbf{Q}}_{i}(\mathbf{Q}^{t},\mathbf{H}^{t}))_{i=1}^{I}$
whose convergence is guaranteed if the stepsizes are properly chosen
{[}cf. Theorem \ref{thm:convergence}{]}. 

Differently from the SISO case, the best-response in (\ref{eq:IC-Q-hat})
does not have a closed-form solution. A standard option to compute
$\hat{\mathbf{Q}}(\mathbf{Q}^{t},\mathbf{H}^{t})$ is using standard
solvers for strongly convex optimization problems. By exploiting the
structure of problem (\ref{eq:IC-update}), we propose next an efficient
iterative algorithm converging to $\hat{\mathbf{Q}}(\mathbf{Q}^{t},\mathbf{H}^{t})$,
wherein the subproblems solved at each step have a closed-form solution. 

\noindent\textbf{Second-order dual method. }To begin with, for notational
simplicity, we rewrite (\ref{eq:IC-Q-hat}) in the following general
form:
\begin{align}
\underset{\mathbf{X}}{\textrm{maximize}}\quad & \begin{array}{l}
\!\!\!\rho\log\det(\mathbf{R}+\mathbf{H}\mathbf{X}\mathbf{H}^{H})+\left<\mathbf{A},\mathbf{X}\right>-\tau\left\Vert \mathbf{X}-\bar{\mathbf{X}}\right\Vert ^{2}\end{array}\nonumber \\
\textrm{subject to}\quad & \mathbf{X}\in\mathcal{Q},\label{eq:logdet-regularized}
\end{align}
where $\mathbf{R}\succ\mathbf{0}$, $\mathbf{A}=\mathbf{A}^{H}$,
$\bar{\mathbf{X}}=\bar{\mathbf{X}}{}^{H}$ and $\mathcal{Q}$ is defined
in (\ref{eq:MIMO-IC-SR}). Let $\mathbf{H}^{H}\mathbf{R}^{-1}\mathbf{H}\triangleq\mathbf{U}\mathbf{D}\mathbf{U}^{H}$
be the eigenvalue/eigenvector decomposition of $\mathbf{H}^{H}\mathbf{R}^{-1}\mathbf{H}$,
where $\mathbf{U}$ is unitary and $\mathbf{D}$ is diagonal with
the diagonal entries arranged in  decreasing order. It is not difficult
to verify that (\ref{eq:logdet-regularized}) is equivalent to the
following problem:
\begin{equation}
\underset{\tilde{\mathbf{X}}\in\mathcal{Q}}{\textrm{maximize}}\quad\rho\log\det(\mathbf{I}+\tilde{\mathbf{X}}\mathbf{D})+\bigl\langle\tilde{\mathbf{A}},\tilde{\mathbf{X}}\bigr\rangle-\tau\bigl\Vert\tilde{\mathbf{X}}-\check{\mathbf{X}}\bigr\Vert^{2},\label{eq:logdet-regularized-s1}
\end{equation}
where $\tilde{\mathbf{X}}\triangleq\mathbf{U}^{H}\mathbf{X}\mathbf{U}$,
$\tilde{\mathbf{A}}\triangleq\mathbf{U}^{H}\mathbf{A}\mathbf{U}$,
and $\check{\mathbf{X}}=\mathbf{U}^{H}\bar{\mathbf{X}}\mathbf{U}$.
We now partition $\mathbf{D}\succeq\mathbf{0}$ in two blocks, its
positive definite and zero parts, and $\tilde{\mathbf{X}}$ accordingly:
\vspace{-0.3cm}

\[
\mathbf{D}=\left[\begin{array}{cc}
\mathbf{D}_{1} & \mathbf{0}\\
\mathbf{0} & \mathbf{0}
\end{array}\right]\quad\mbox{and}\quad\mathbf{\tilde{\mathbf{X}}}=\left[\begin{array}{cc}
\tilde{\mathbf{X}}_{11} & \tilde{\mathbf{X}}_{12}\\
\tilde{\mathbf{X}}_{21} & \tilde{\mathbf{X}}_{22}
\end{array}\right]
\]
where $\mathbf{D}_{1}\succ\mathbf{0}$, and $\tilde{\mathbf{X}}_{11}$
and $\mathbf{D}_{1}$ have the same dimensions. Problem (\ref{eq:logdet-regularized-s1})
can be then rewritten as:
\begin{equation}
\underset{\tilde{\mathbf{X}}\in\mathcal{Q}}{\textrm{maximize}}\quad\rho\log\det(\mathbf{I}+\tilde{\mathbf{X}}_{11}\mathbf{D}_{1})+\bigl\langle\tilde{\mathbf{A}},\tilde{\mathbf{X}}\bigr\rangle-\tau\bigl\Vert\tilde{\mathbf{X}}-\check{\mathbf{X}}\bigr\Vert^{2},\label{eq:logdet-regularized-s2}
\end{equation}
 Note that, since $\tilde{\mathbf{X}}\in\mathcal{Q}$, by definition
$\tilde{\mathbf{X}}_{11}$ must belong to $\mathcal{Q}$ as well.
Using this fact and introducing the slack variable $\mathbf{Y}=\tilde{\mathbf{X}}_{11}$,
(\ref{eq:logdet-regularized-s2}) is equivalent to
\begin{align}
\underset{\tilde{\mathbf{X}},\mathbf{Y}}{\textrm{maximize}}\quad & \rho\log\det(\mathbf{I}+\mathbf{Y}\mathbf{D}_{1})+\bigl\langle\tilde{\mathbf{A}},\tilde{\mathbf{X}}\bigr\rangle-\tau\bigl\Vert\tilde{\mathbf{X}}-\check{\mathbf{X}}\bigr\Vert^{2}\nonumber \\
\textrm{subject to}\quad & \tilde{\mathbf{X}}\in\mathcal{Q},\,\mathbf{Y}=\tilde{\mathbf{X}}_{11},\,\mathbf{Y}\in\mathcal{Q}.\label{eq:logdet-regularized-s3}
\end{align}

Now we solve (\ref{eq:logdet-regularized-s3}) via dual decomposition
(note that there is zero duality gap). The (partial) Lagrangian function
of (\ref{eq:logdet-regularized-s3}) is: denoting by $\mathbf{Z}$
the matrix of multipliers associated to the linear constraints $\mathbf{Y}=\tilde{\mathbf{X}}_{11}$,
\[
\begin{split}L(\tilde{\mathbf{X}},\mathbf{Y},\mathbf{Z})=\; & \rho\log\det(\mathbf{I}+\mathbf{Y}\mathbf{D}_{1})+\bigl\langle\tilde{\mathbf{A}},\tilde{\mathbf{X}}\bigr\rangle\\
 & -\tau\bigl\Vert\tilde{\mathbf{X}}-\check{\mathbf{X}}\bigr\Vert^{2}+\bigl\langle\mathbf{Z},\mathbf{Y}-\tilde{\mathbf{X}}_{11}\bigr\rangle.
\end{split}
\]
The dual problem is then 
\[
\underset{\mathbf{Z}}{\textrm{minimize}}\quad d(\mathbf{Z})=L(\tilde{\mathbf{X}}(\mathbf{Z}),\mathbf{Y}(\mathbf{Z}),\mathbf{Z}),
\]
with
\begin{align}
\tilde{\mathbf{X}}(\mathbf{Z}) & =\underset{\tilde{\mathbf{X}}\in\mathcal{Q}}{\arg\max}\;-\tau\bigl\Vert\tilde{\mathbf{X}}-\check{\mathbf{X}}\bigr\Vert^{2}-\bigl\langle\mathbf{Z},\tilde{\mathbf{X}}_{11}\bigr\rangle,\label{eq:logdet-regularized-s4-1}\\
\mathbf{Y}(\mathbf{Z}) & =\underset{\mathbf{Y}\in\mathcal{Q}}{\arg\max}\;\rho\log\det(\mathbf{I}+\mathbf{Y}\mathbf{D}_{1})+\left\langle \mathbf{Z},\mathbf{Y}\right\rangle .\label{eq:logdet-regularized-s4-2}
\end{align}

Problem (\ref{eq:logdet-regularized-s4-1}) is quadratic and has a
closed-form solution (see Lemma \ref{lem:waterfilling-solution} below).
Similarly, if $\mathbf{Z}\prec\mathbf{0}$, (\ref{eq:logdet-regularized-s4-2})
can be solved in closed-form, up to a Lagrange multiplier which can
be found by bisection; see, e.g., \cite[Table I]{Kim2011}. In our
setting, however, $\mathbf{Z}$ in (\ref{eq:logdet-regularized-s4-2})
is not necessarily negative definite. Nevertheless, the next lemma
provides a closed form expression of $\mathbf{Y(Z)}$ {[}and $\tilde{\mathbf{X}}(\mathbf{Z})${]}. 
\begin{lem}
\label{lem:waterfilling-solution}Given (\ref{eq:logdet-regularized-s4-1})
and (\ref{eq:logdet-regularized-s4-2}) in the setting above, the
following hold: \end{lem}
\begin{description}
\item [{i)}] $\tilde{\mathbf{X}}(\mathbf{Z})$ in (\ref{eq:logdet-regularized-s4-1})
is given by 
\begin{equation}
\tilde{\mathbf{X}}(\mathbf{Z})=\left[\check{\mathbf{X}}-\frac{1}{2\tau}\left(\mu^{\star}\mathbf{I}+\left[\begin{array}{cc}
\mathbf{Z} & \mathbf{0}\\
\mathbf{0} & \mathbf{0}
\end{array}\right]\right)\right]^{+},\label{eq:X-closed_form}
\end{equation}
where $[\mathbf{X}]^{+}$ denotes the projection of $\mathbf{X}$
onto the cone of  positive semidefinite matrices, and $\mu^{\star}$
is the multiplier such that $0\leq\mu^{\star}\perp\textrm{tr}(\tilde{\mathbf{X}}(\mathbf{Z}))-P\leq0$,
which can be found by bisection; 
\item [{ii)}] $\mathbf{Y(Z)}$ in (\ref{eq:logdet-regularized-s4-2}) is
unique and is given by
\begin{equation}
\mathbf{Y(Z)}=\mathbf{V}\,[\rho\,\mathbf{I}-\boldsymbol{\Sigma}^{-1}]^{+}\,\mathbf{V}^{H},\label{eq:Y_closed_form}
\end{equation}
where $(\mathbf{V},\boldsymbol{\Sigma})$ is the generalized eigenvalue
decomposition of $(\mathbf{D}_{1},-\mathbf{Z}+\mu^{\star}\mathbf{I})$,
and $\mu^{\star}$ is the multiplier such that $0\leq\mu^{\star}\perp\textrm{tr}(\mathbf{Y(Z)})-P\leq0$;
$\mu^{\star}$ can be found by bisection over $[\underline{\mu},\overline{\mu}]$,
with $\underline{\mu}\triangleq[\lambda_{\max}(\mathbf{Z})]^{+}$
and $\overline{\mu}\triangleq[\lambda_{\max}(\mathbf{D}_{1})+\lambda_{\max}(\mathbf{Z})/\rho]^{+}.$
\end{description}
\begin{proof} See Appendix \ref{sec:Proof-of-waterfilling-solution}.\end{proof}\smallskip 

Since $(\tilde{\mathbf{X}}(\mathbf{Z}),\mathbf{Y}(\mathbf{Z}))$ is
unique, $d(\mathbf{Z})$ is differentiable \nocite{bertsekas03},
with conjugate gradient \cite{Scutari2012a} 
\[
\nabla_{\mathbf{Z}^{*}}d(\mathbf{Z})=\mathbf{Y(Z)}-\tilde{\mathbf{X}}_{11}(\mathbf{Z}).
\]
 One can then solve the dual problem using standard (proximal) gradient-based
methods; see, e.g., \cite{bertsekas1999nonlinear}. As a matter of
fact, $d(\mathbf{Z})$ is twice continuously differentiable, whose
augmented Hessian matrix \cite{Scutari2012a} is given by \cite[Sec. 4.2.4]{bertsekas1999nonlinear}:
\[
\begin{array}{l}
\nabla_{\mathbf{Z}\mathbf{Z}^{*}}^{2}d(\mathbf{Z})=\;-\left[\mathbf{I}\;-\mathbf{I}\right]^{H}\cdot\smallskip\\
\qquad\Bigl[\textrm{bdiag}(\nabla_{\mathbf{Y}\mathbf{Y}^{*}}^{2}L(\tilde{\mathbf{X}},\mathbf{Y,Z}),\nabla_{\tilde{\mathbf{X}}_{11}\tilde{\mathbf{X}}_{11}^{*}}^{2}L(\tilde{\mathbf{X}},\mathbf{Y,Z}))\Bigr]^{-1}\cdot\smallskip\\
\qquad\left[\mathbf{I}\;-\mathbf{I}\right]\bigr|_{\tilde{\mathbf{X}}=\tilde{\mathbf{X}}\mathbf{(Z)},\mathbf{Y=Y(Z)}},
\end{array}
\]
with 
\[
\begin{split}\nabla_{\mathbf{Y}\mathbf{Y}^{*}}^{2}L(\tilde{\mathbf{X}},\mathbf{Y,Z}) & =-\rho^{2}\cdot(\mathbf{D}_{1}^{1/2}(\mathbf{I}+\mathbf{D}_{1}^{1/2}\mathbf{Y}\mathbf{D}_{1}^{1/2})^{-1}\mathbf{D}_{1}^{1/2})^{T}\\
 & \quad\otimes(\mathbf{D}_{1}^{1/2}(\mathbf{I}+\mathbf{D}_{1}^{1/2}\mathbf{Y}\mathbf{D}_{1}^{1/2})^{-1}\mathbf{D}_{1}^{1/2}),
\end{split}
\]
and $\nabla_{\tilde{\mathbf{X}}_{11}\tilde{\mathbf{X}}_{11}^{*}}^{2}L(\tilde{\mathbf{X}}\mathbf{Y,Z})=-\tau\mathbf{I}$.
Since $\mathbf{D}_{1}\succ\mathbf{0}$, it is easy to verify that
$\nabla_{\mathbf{Z}\mathbf{Z}^{*}}^{2}d(\mathbf{Z})\succ\mathbf{0}$
and the following second-order Newton's method to update the dual
variable $\mathbf{Z}$ is well-defined:
\[
\textrm{vec}(\mathbf{Z}^{t+1})=\textrm{vec}(\mathbf{Z}^{t})-(\nabla_{\mathbf{Z}\mathbf{Z}^{*}}^{2}d(\mathbf{Z}^{t}))^{-1}\textrm{vec}(\nabla d(\mathbf{Z}^{t})).
\]
The convergence speed of the Newton's methods is typically very fast,
and, in particular, superlinear convergence rate can be expected when
$\mathbf{Z}^{t}$ is close to\textbf{ $\mathbf{Z}^{\star}$ }\cite[Prop. 1.4.1]{bertsekas1999nonlinear}.\qed\smallskip

\begin{figure*}[t]
\center\subfigure[ergodic sum-rate versus iterations]{\includegraphics[scale=0.6]{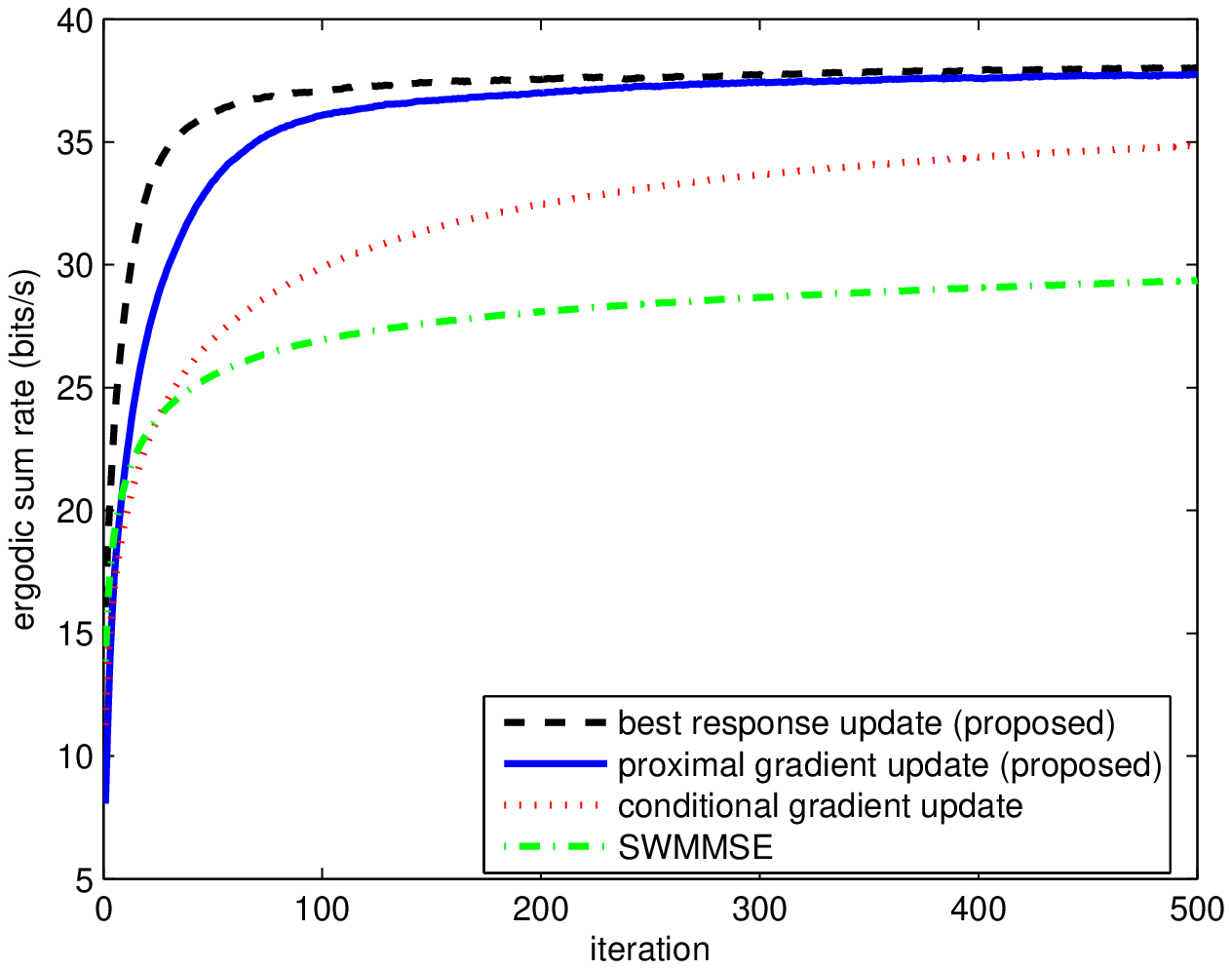}}\hspace{0.9cm}\subfigure[achievable sum-rate versus iterations]{\includegraphics[scale=0.6]{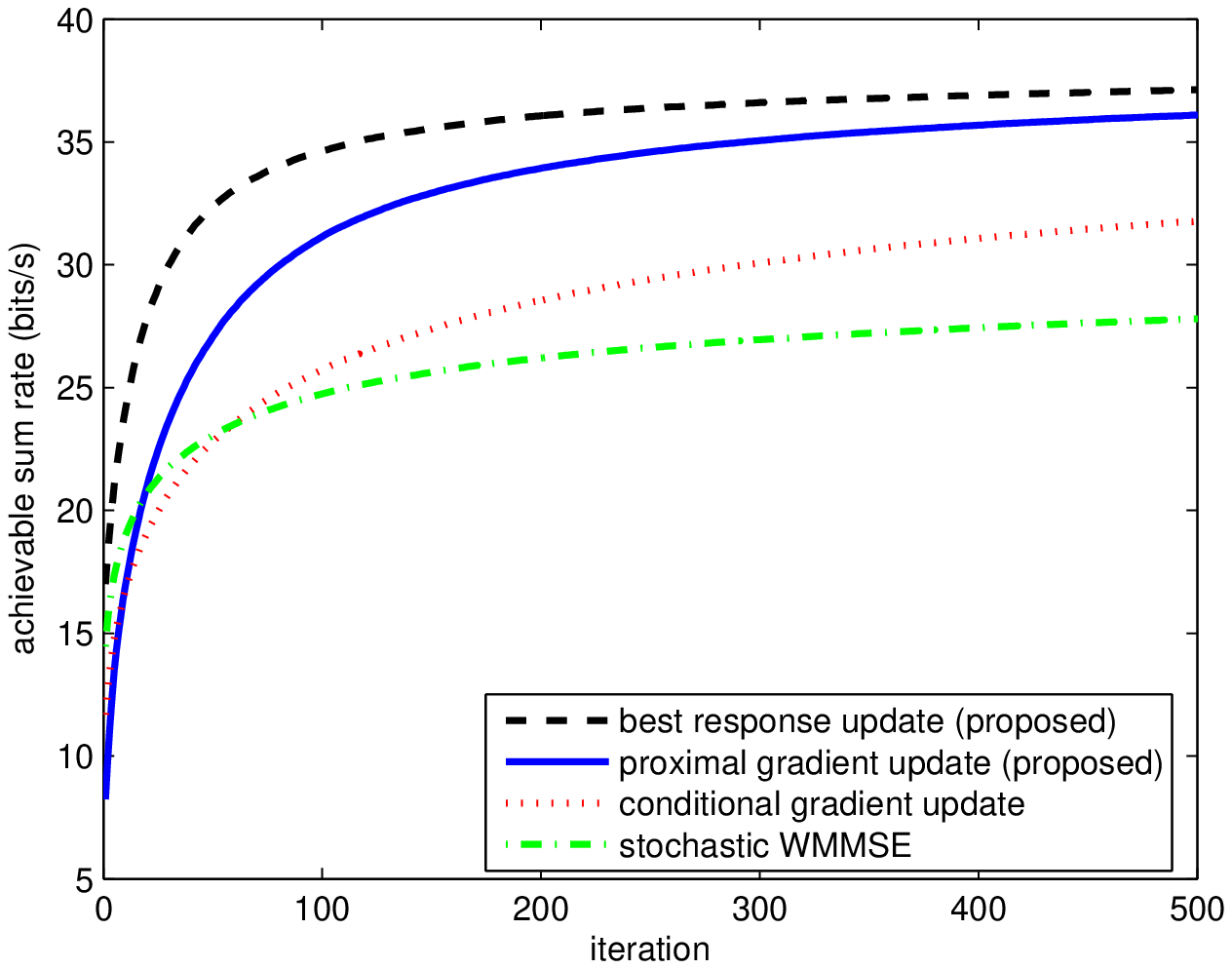}}\protect\caption{\label{fig:MIMO-IC-SR}Sum-rate versus iteration in a 50-user MIMO
IC}
\end{figure*}

As a final remark on efficient solution methods computing $\hat{\mathbf{Q}}_{i}(\mathbf{Q}^{t},\mathbf{H}^{t})$,
note that one can also apply the \emph{proximal} conditional gradient
method as introduced in (\ref{eq:conditional-gradient}), which is
based on a fully linearization of the social function plus a proximal
regularization term:
\begin{equation}
\begin{split}\hat{\mathbf{Q}}_{i}(\mathbf{Q}^{t},\mathbf{H}^{t}) & =\underset{\mathbf{Q}_{i}\in\mathcal{Q}}{\arg\max}\Bigl\{\bigl\langle\mathbf{Q}_{i}-\mathbf{Q}_{i}^{t},\mathbf{F}_{i}^{t}\bigr\rangle-\tau_{i}\,\bigl\Vert\mathbf{Q}_{i}-\mathbf{Q}_{i}^{t}\bigr\Vert^{2}\Bigr\}\\
 & =\left[\mathbf{Q}_{i}^{t}+\frac{1}{2\tau_{i}}(\mathbf{F}_{i}^{t}-\mu^{\star}\mathbf{I})\right]^{+},
\end{split}
\raisetag{\baselineskip}\label{eq:IC-Proximal-Gradient}
\end{equation}
where $\mu^{\star}$ is the Lagrange multiplier that can be found
efficiently by bisection method. Note that (\ref{eq:IC-Proximal-Gradient})
differs from more traditional conditional stochastic gradient methods
\cite{Ermol'ev1977} by the presence of the proximal regularization,
thanks to which one can solve (\ref{eq:IC-Proximal-Gradient}) in
closed form. 

\noindent\textbf{Practical implementations}\emph{. }The proposed
algorithm is fairly distributed: once the pricing matrix $\boldsymbol{\Pi}_{i}$
is given, to compute the best-response, each user only needs to locally
estimate the covariance matrix of the interference plus noise. Note
that both the computation of $\hat{\mathbf{Q}}_{i}(\mathbf{Q},\mathbf{H})$
and the update of $\mathbf{F}_{i}$ can be implemented locally by
each user. The estimation of the pricing matrix $\boldsymbol{\Pi}_{i}$
requires however some signaling among nearby receivers. Quite interestingly,
the pricing expression and thus the resulting signaling overhead necessary
to compute it coincide with \cite{Kim2011} (where a sequential algorithm
is proposed for the deterministic maximization of the sum-rate over
MIMO ICs) and the stochastic gradient projection method in \cite{DiLorenzo2013}.
We remark that the signaling to compute (\ref{eq:IC-pricing}) is
lower than in \cite{Razaviyayn_spawc2013,Razaviyayn_2013}, wherein
signaling exchange is required twice (one in the computation of $\mathbf{U}_{i}$
and another in that of $\mathbf{A}_{i}$; see \cite[Algorithm 1]{Razaviyayn_spawc2013}
for more details) in a single iteration to transmit among users the
auxiliary variables which are of same dimensions as $\boldsymbol{\Pi}_{i}$.

\noindent\textbf{Numerical Results}\emph{.} We considered the same
scenario as in the SISO case (cf. Sec. \ref{sub:SISO-IC}) with the
following differences: i) there are 50 users; ii) the channels are
matrices generated according to $\mathbf{H}^{t}=\mathbf{H}+\triangle\mathbf{H}^{t}$,
where $\mathbf{H}$ is given while $\triangle\mathbf{H}^{t}$ is realization
dependent and generated by $\delta\cdot\texttt{randn}$, with noise
level $\delta=0.2$; and iii) the number of transmit and receive antennas
is four. We simulate the following algorithms: i) the proposed stochastic
best-response pricing algorithm (\ref{eq:IC-update}) (with $\tau_{i}=10^{-8}$
for all $i$; $\gamma^{1}=\rho^{0}=\rho^{1}=1$ and $\rho^{t}=2/(t+2)^{0.6}$
and $\gamma^{t}=2/(t+2)^{0.61}$ for $t\geq2$); ii) the proposed
stochastic proximal gradient method (\ref{eq:IC-Proximal-Gradient})
with $\tau=0.01$ and same stepsize as stochastic best-response pricing
algorithm; iii) the stochastic conditional gradient method \cite{Ermol'ev1977}
(with $\gamma^{1}=\rho^{0}=\rho^{1}=1$ and $\rho^{t}=1/(t+2)^{0.9}$
and $\gamma^{t}=1/(t+2)^{0.91}$ for $t\geq2$); and iv) the stochastic
weighted minimum mean-square-error (SWMMSE) method \cite{Razaviyayn_spawc2013}.
The best-response of the algorithm in i) is computed using Lemma \ref{lem:waterfilling-solution}.
We observed convergence of the inner loop solving (\ref{eq:logdet-regularized-s4-1})-(\ref{eq:logdet-regularized-s4-2})
in a very few iterations. Similarly to the SISO ICs case, we consider
both ergodic sum-rate and achievable sum-rate. In Figure \ref{fig:MIMO-IC-SR}
we plot both objective functions versus the iteration index. It is
clear from the figures that the proposed best-response pricing and
proximal gradient algorithms outperform current schemes in terms of
both convergence speed and achievable (ergodic or instantaneous) sum-rate.
Note also that the best-response pricing algorithm is very scalable
compared with the other algorithms. Finally, it is interesting to
note that  the proposed stochastic proximal gradient algorithm outperforms
the conditional stochastic gradient method in terms of both convergence
speed and iteration complexity. This is mainly due to the presence
of the proximal regularization. \vspace{-0.2cm}

\subsection{\label{sub:MIMO-MAC}Sum-rate maximization over MIMO MACs}

\begin{figure*}[tbh]
 \center\subfigure[ergodic sum-rate versus iterations]{\includegraphics[bb=115bp 277bp 472bp 554bp,clip,scale=0.6]{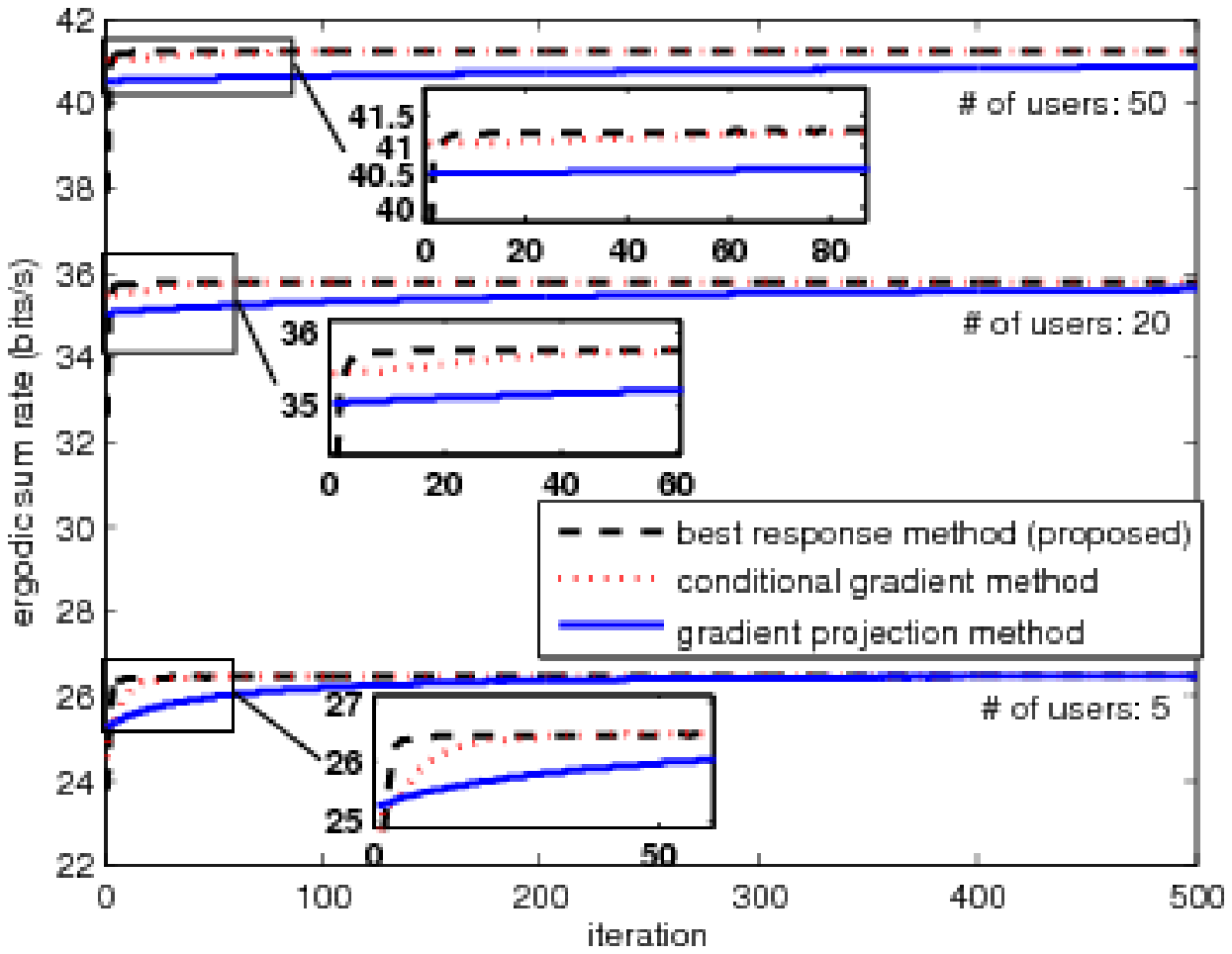}}\hspace{0.9cm}\subfigure[achievable sum-rate versus iterations]{\includegraphics[bb=115bp 277bp 472bp 554bp,clip,scale=0.6]{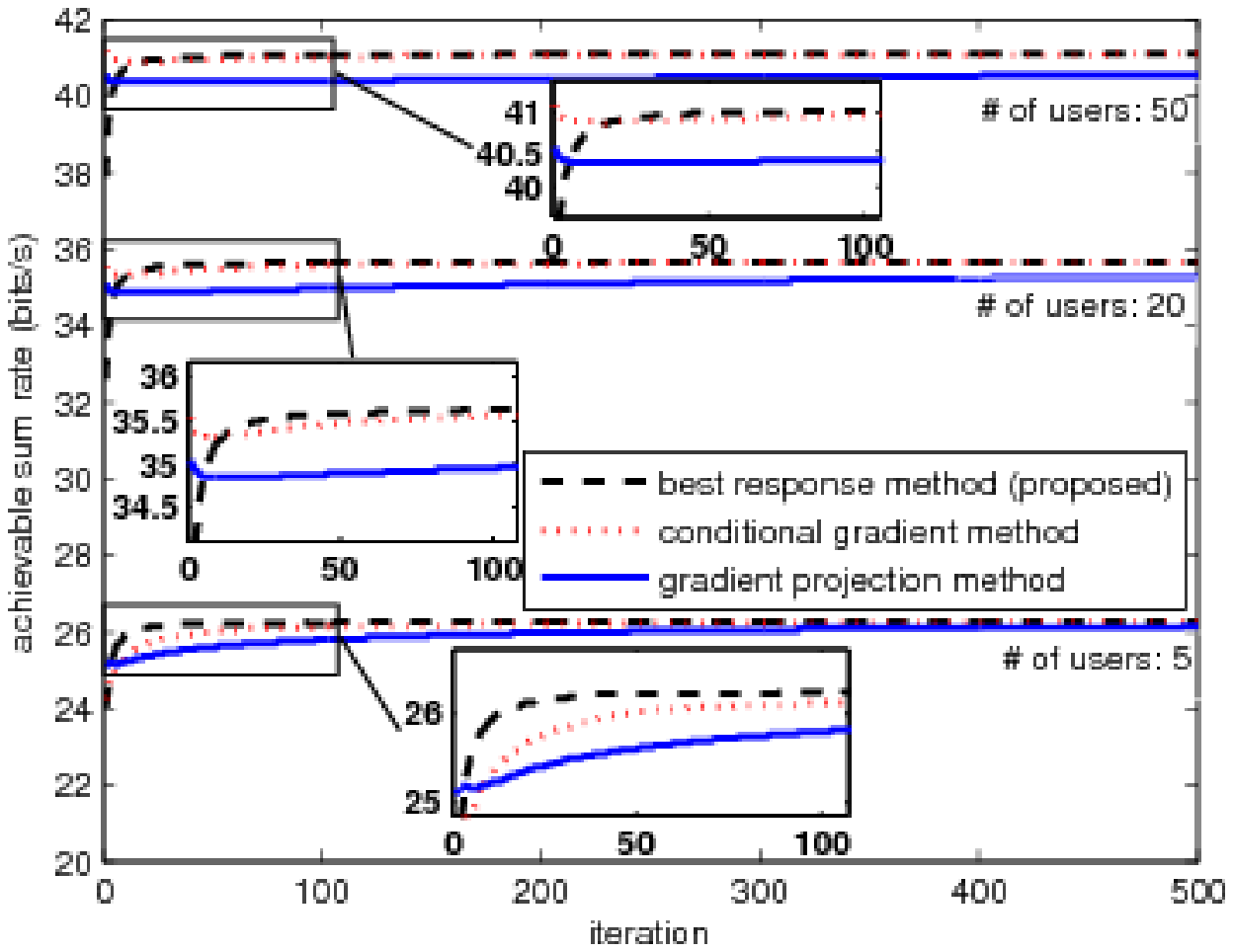}}\protect\caption{\label{fig:MIMO-MAC-SR}Sum-rate versus iteration in MIMO MAC}
\vspace{-0.5cm}
\end{figure*}

We consider now the sum-rate maximization problem over MIMO MACs,
as defined in (\ref{eq:MIMO-MAC-SR}). Define
\[
r(\mathbf{H},\mathbf{Q})\triangleq\log\det\left(\mathbf{R}_{\textrm{N}}+{\textstyle \sum_{i=1}^{I}}\mathbf{H}_{i}\mathbf{Q}_{i}\mathbf{H}_{i}^{H}\right).
\]
A natural choice for the best-response of each user $i$ is {[}cf.
(\ref{eq:convex-jacobi}){]}:
\begin{align}
\hat{\mathbf{Q}}_{i}(\mathbf{Q}^{t},\mathbf{H}^{t}) & =\underset{\mathbf{Q}_{i}\in\mathcal{Q}_{i}}{\arg\max}\Bigl\{\rho^{t}\, r(\mathbf{H}^{t},\mathbf{Q}_{i},\mathbf{Q}_{-i}^{t})\nonumber \\
+(1-\rho^{t}) & \bigl\langle\mathbf{Q}_{i}-\mathbf{Q}_{i}^{t},\mathbf{F}_{i}^{t-1}\bigr\rangle-\tau_{i}\left\Vert \mathbf{Q}_{i}-\mathbf{Q}_{i}^{t}\right\Vert ^{2}\Bigr\},\label{eq:MAC-Q-hat}
\end{align}
 and $\mathbf{F}_{i}^{t}$ is updated as $\mathbf{F}_{i}^{t}=(1-\rho^{t})\mathbf{F}_{i}^{t-1}+\rho^{t}\,\nabla_{\mathbf{Q}_{i}^{*}}r(\mathbf{H}^{t},\mathbf{Q}^{t})$
while $\nabla_{\mathbf{Q}_{i}^{*}}r(\mathbf{H},\mathbf{Q})=\mathbf{H}_{i}^{H}(\mathbf{R}_{\textrm{N}}+\sum_{i=1}^{I}\mathbf{H}_{i}\mathbf{Q}_{i}\mathbf{H}_{i}^{H})^{-1}\mathbf{H}_{i}$.

Note that since the instantaneous sum-rate function $\log\det(\mathbf{R}_{\textrm{N}}+\sum_{i=1}^{I}\mathbf{H}_{i}\mathbf{Q}_{i}\mathbf{H}^{H})$
is jointly concave in $\mathbf{Q}_{i}$ for any $\mathbf{H}$, the
ergodic sum-rate function is concave in $\mathbf{Q}_{i}$'s, and thus
Algorithm \ref{alg} will converge (in the sense of Theorem 1) to
the \emph{global }optimal solution of (\ref{eq:MIMO-MAC-SR}). To
the best of our knowledge, this is the first example of stochastic
approximation algorithm based on best-response dynamics rather than
gradient responses. 

\noindent\textbf{Numerical results.}\emph{ }We compare the proposed
best-response method (\ref{eq:MAC-Q-hat}) (whose solution is computed
using Method \#2 in Sec. \ref{sub:application:MIMO-IC}) with the
stochastic conditional gradient method \cite{Ermol'ev1977}, and the
stochastic gradient projection method \cite{Yousefian2012}. System
parameters (including the stepsize rules) are set as for the MIMO
IC example in Sec. \ref{sub:application:MIMO-IC}. In Figure \ref{fig:MIMO-MAC-SR}
we plot both the ergodic sum-rate and the achievable sum-rate versus
the iteration index. The figure clearly shows that Algorithm \ref{alg}
outperforms the conditional gradient method and the gradient projection
method in terms of convergence speed, and the performance gap is increasing
as the number of users increases. This is because the proposed algorithm
is a best-response type scheme, which thus explores the concavity
of each users' rate function better than what gradient methods do.
Note also the good scalability of the proposed best-response method.\vspace{-0.2cm}

\subsection{Distributed deterministic algorithms with errors}

The developed framework can be useful in the context of deterministic
optimization as well to robustify best-response-based algorithms in
the presence of noisy estimates of system parameters. Consider the
\emph{deterministic} optimization problem introduced in (\ref{eq:deterministic_social}).
The main iterate of the deterministic counterpart of Algorithm \ref{alg}
is still given by (\ref{eq:update-rule-x}) but with each $\widehat{\mathbf{x}}_{i}(\mathbf{x}^{t})$
defined as \cite{Scutarib}\vspace{-0.1cm}
\begin{equation}
\widehat{\mathbf{x}}_{i}(\mathbf{x}^{t})=\underset{\mathbf{x}_{i}\in\mathcal{X}_{i}}{\arg\min}\Biggl\{\!\!\!\begin{array}{l}
{\textstyle \sum_{j\in\mathcal{C}_{i}}}f_{i}(\mathbf{x}_{i},\mathbf{x}_{-i}^{t})+\bigl\langle\mathbf{x}_{i}-\mathbf{x}_{i}^{t},\boldsymbol{\pi}_{i}(\mathbf{x}^{t})\bigr\rangle\smallskip\\
+\tau_{i}\left\Vert \mathbf{x}_{i}-\mathbf{x}_{i}^{t}\right\Vert ^{2},
\end{array}\!\!\!\Biggr\},\label{eq:deterministic-pricing}
\end{equation}
 where $\boldsymbol{\pi}_{i}(\mathbf{x})=\sum_{j\in\overline{\mathcal{C}}_{i}}\nabla_{i}f_{j}(\mathbf{x})$.
In many applications (see, e.g., \cite{Zhang2008b,Hong2011,DiLorenzo2013}),
however, only a noisy estimate, denoted by $\widetilde{\boldsymbol{\pi}}_{i}(\mathbf{x})$,
is available instead of $\boldsymbol{\pi}_{i}(\mathbf{x})$. A heuristic
is then to replace in (\ref{eq:deterministic-pricing}) the exact
$\boldsymbol{\pi}_{i}(\mathbf{x})$ with its noisy estimate $\widetilde{\boldsymbol{\pi}}_{i}(\mathbf{x})$.
The limitation of this approach, albeit natural, is that convergence
of the resulting scheme is in jeopardy. 

If $\widetilde{\boldsymbol{\pi}}_{i}(\mathbf{x})$ is unbiased, i.e.,
$\mathbb{E}\left[\widetilde{\boldsymbol{\pi}}_{i}(\mathbf{x}^{t})|\mathcal{F}^{t}\right]=\boldsymbol{\pi}_{i}(\mathbf{x}^{t})$
\cite{Zhang2008b,Hong2011}, capitalizing on the proposed framework,
we can readily deal with estimation errors while guaranteeing convergence.
In particular, it is sufficient to modify (\ref{eq:deterministic-pricing})
as follows:
\begin{equation}
\begin{split}\widetilde{\mathbf{x}}_{i}(\mathbf{x}^{t})=\; & \underset{\mathbf{x}_{i}\in\mathcal{X}_{i}}{\arg\min}\;\Bigl\{{\textstyle \sum_{j\in\mathcal{C}_{i}}}f_{j}(\mathbf{x}_{i},\mathbf{x}_{-i}^{t})+\rho^{t}\bigl\langle\mathbf{x}_{i}-\mathbf{x}_{i}^{t},\widetilde{\boldsymbol{\pi}}_{i}(\mathbf{x}^{t})\bigr\rangle\\
+ & (1-\rho^{t})\bigl\langle\mathbf{x}_{i}-\mathbf{x}_{i}^{t},\mathbf{f}_{i}^{t-1}\bigr\rangle+\tau_{i}\left\Vert \mathbf{x}_{i}-\mathbf{x}_{i}^{t}\right\Vert ^{2}\Bigr\},
\end{split}
\raisetag{\baselineskip}\label{eq:deterministic-robust-update-1}
\end{equation}
where $\mathbf{f}_{i}^{t}$ is updated according to $\mathbf{f}_{i}^{t}=(1-\rho^{t})\mathbf{f}_{i}^{t-1}+\rho^{t}\widetilde{\boldsymbol{\pi}}_{i}(\mathbf{x}^{t}).$
Algorithm \ref{alg} based on the best-response (\ref{eq:deterministic-robust-update-1})
is then guaranteed to converge to a stationary solution of (\ref{eq:deterministic_social}),
in the sense specified by Theorem \ref{thm:convergence}.

As a case study, we consider next the maximization of the deterministic
sum-rate over MIMO ICs in the presence of pricing estimation errors:
\begin{align}
\underset{\mathbf{Q}}{\textrm{maximize}}\quad & {\textstyle \sum_{i=1}^{I}}\log\det(\mathbf{I}+\mathbf{H}_{ii}\mathbf{Q}_{i}\mathbf{H}_{ii}^{H}\mathbf{R}_{i}(\mathbf{Q}_{-i})^{-1})\nonumber \\
\textrm{subject to}\quad & \mathbf{Q}_{i}\succeq\mathbf{0},\,\textrm{tr}(\mathbf{Q}_{i})\leq P_{i},\quad i=1,\ldots,I.\label{eq:MIMO-IC-deterministic}
\end{align}
Then (\ref{eq:deterministic-robust-update-1}) becomes:
\begin{equation}
\begin{array}{l}
\widehat{\mathbf{Q}}_{i}(\mathbf{Q}^{t})=\underset{\mathbf{Q}_{i}\in\mathcal{Q}_{i}}{\arg\max}\Bigl\{\log\det\left(\mathbf{R}_{i}^{t}+\mathbf{H}_{ii}^{t}\mathbf{Q}_{i}(\mathbf{H}_{ii}^{t})^{H}\right)\smallskip\\
\quad+\bigl\langle\mathbf{Q}_{i}-\mathbf{Q}_{i}^{t},\rho^{t}\widetilde{\boldsymbol{\Pi}}_{i}^{t}+(1-\rho^{t})\mathbf{F}_{i}^{t-1}\bigr\rangle-\tau_{i}\bigl\Vert\mathbf{Q}_{i}-\mathbf{Q}_{i}^{t}\bigr\Vert^{2}\Bigr\},
\end{array}\label{eq:MIMO-IC-deterministic-update-1}
\end{equation}
where $\widetilde{\boldsymbol{\Pi}}_{i}^{t}$ is a noisy estimate
of $\boldsymbol{\Pi}_{i}(\mathbf{Q}^{t},\mathbf{H})$ given by (\ref{eq:IC-pricing})%
\footnote{\textbf{$\boldsymbol{\Pi}_{i}(\mathbf{Q},\mathbf{H})$} is always
negative definite by definition \cite{Kim2011}, but $\widetilde{\boldsymbol{\Pi}}_{i}^{t}$
may not be so. However, it is reasonable to assume $\widetilde{\boldsymbol{\Pi}}_{i}^{t}$
to be Hermitian.%
} and $\mathbf{F}_{i}^{t}$ is updated according to $\mathbf{F}_{i}^{t}=\rho^{t}\widetilde{\boldsymbol{\Pi}}_{i}^{t}+(1-\rho^{t})\mathbf{F}_{i}^{t-1}.$
Given $\widehat{\mathbf{Q}}_{i}(\mathbf{Q}^{t})$, the main iterate
of the algorithm becomes $\mathbf{Q}_{i}^{t+1}=\mathbf{Q}_{i}^{t}+\gamma^{t+1}\left(\hat{\mathbf{Q}}_{i}(\mathbf{Q}^{t})-\mathbf{Q}_{i}^{t}\right).$
Convergence w.p.1 to a stationary point of the deterministic optimization
problem (\ref{eq:MIMO-IC-deterministic}) is guaranteed by Theorem
\ref{thm:convergence}. Note that if the channel matrices $\{\mathbf{H}_{ii}\}$
are full column-rank, one can also set in (\ref{eq:MIMO-IC-deterministic-update-1})
all $\tau_{i}=0$, and compute (\ref{eq:MIMO-IC-deterministic-update-1})
in closed form (cf. Lemma \ref{lem:waterfilling-solution}).

\begin{figure}[t]
\center\includegraphics[scale=0.6]{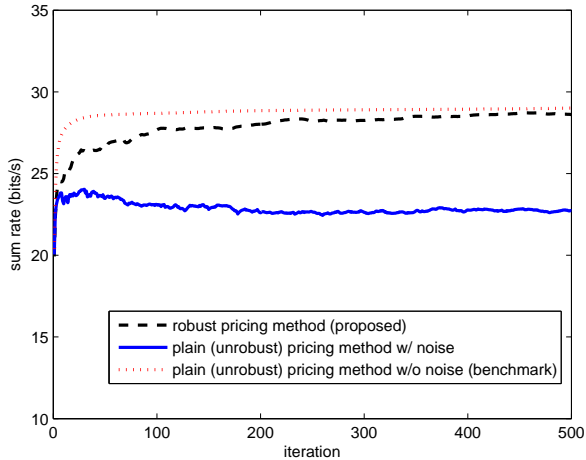}\protect\caption{\label{fig:MIMO-IC-SR-noisy-parameters}Maximization of deterministic
sum-rate over MIMO IC under noisy parameter estimation: sum-rate versus
iteration. }

\vspace{-0.8cm}
\end{figure}

\noindent\textbf{Numerical results}\emph{. }We consider the maximization
of the deterministic sum-rate (\ref{eq:MIMO-IC-deterministic}) over
a 5-user MIMO IC. The other system parameters (including the stepsize
rules) are set as in the numerical example in Sec. \ref{sub:application:MIMO-IC}.
The noisy estimate $\widetilde{\boldsymbol{\Pi}}_{i}$ of the nominal
price matrix $\boldsymbol{\Pi}_{i}$ {[}defined in (\ref{eq:IC-pricing}){]}
is $\widetilde{\boldsymbol{\Pi}}_{i}^{t}=\boldsymbol{\Pi}_{i}+\Delta\boldsymbol{\Pi}_{i}^{t}$,
where $\Delta\boldsymbol{\Pi}_{i}^{t}$ is firstly generated as $\Delta\mathbf{H}^{t}$
in Sec. \ref{sub:application:MIMO-IC} and then only its Hermitian
part is kept; the noise level $\delta$ is set to 0.05. We compare
the following algorithms: i) the proposed robust pricing method$-$Algorithm
1 based on the best-response defined in (\ref{eq:MIMO-IC-deterministic-update-1});
and ii) the plain pricing method as proposed in \cite{Scutarib} {[}cf.
(\ref{eq:deterministic-pricing}){]}. We also include as a benchmark
the sum-rate achieved by the plain pricing method (\ref{eq:deterministic-pricing})
when there is no estimation noise (i.e., perfect $\boldsymbol{\pi}_{i}(\mathbf{x})$
is available). In Figure \ref{fig:MIMO-IC-SR-noisy-parameters} we
plot the deterministic sum-rate in (\ref{eq:MIMO-IC-deterministic})
versus the iteration index $t$. As expected, Figure \ref{fig:MIMO-IC-SR-noisy-parameters}
shows that the plain pricing method \cite{Scutarib} is not robust
to pricing estimation errors, whereas the proposed robustification
preforms quite well. For instance, the rate achievable by the proposed
method is about 50\% larger than the one of \cite{Scutarib}, and
is observed to reach the benchmark value (achieved by the plain pricing
method when there is no estimation noise). This is due to the fact
that the proposed robustification filters out the estimation noise.
Note that the limit point generated by the proposed scheme is a stationary
solution of the deterministic problem (\ref{eq:MIMO-IC-deterministic}).

\section{Conclusions\label{sec:Conclusions}}

In this paper, we have proposed a novel best-response-based algorithmic
framework converging to stationary solutions of general stochastic
nonconvex optimization problems. The proposed novel decomposition
enables all users to update their optimization variables \emph{in
parallel} by solving a sequence of strongly convex subproblems; which
makes the algorithm very appealing for the distributed design of several
multi-agent systems. We have then customized the general framework
to solve special classes of problems as well as specific applications,
including the stochastic maximization of the sum-rate over frequency-selective
ICs, MIMO ICs and MACs. Extensive experiments have provided a solid
evidence of the superiority in terms of both achievable sum-rate and
practical convergence of the proposed schemes with respect to to state-of-the-art
stochastic-based algorithms.

\appendix

\subsection{\label{sec:Proof-of-convergence}Proof of Theorem \ref{thm:convergence}}

We first introduce the following two preliminary results. 
\begin{lem}
\label{lem:incremental-gradient}Given problem (\ref{eq:social-formulation})
under Assumptions (a)-(c), suppose that the stepsizes $\{\gamma^{t}\}$
and $\{\rho^{t}\}$ are chosen according to (\ref{eq:stepsize}).
Let $\{\mathbf{x}^{t}\}$ be the sequence generated by Algorithm 1.
Then, the following holds 
\[
\lim_{t\rightarrow\infty}\left\Vert \mathbf{f}^{t}-\nabla U(\mathbf{x}^{t})\right\Vert =0,\qquad\mbox{\emph{w.p.1}}.
\]
\end{lem}
\begin{IEEEproof}
This lemma is a consequence of \cite[Lemma 1]{Ruszczynski1980}. To
see this, we just need to verify that all the technical conditions
therein are satisfied by the problem at hand. Specifically, Condition
(a) of \cite[Lemma 1]{Ruszczynski1980} is satisfied because $\mathcal{X}_{i}$'s
are closed and bounded in view of Assumption (a). Condition (b) of
\cite[Lemma 1]{Ruszczynski1980} is exactly Assumption (c). Conditions
(c)-(d) come from the stepsize rules i)-ii) in (\ref{eq:stepsize})
of Theorem \ref{thm:convergence}. Condition (e) of \cite[Lemma 1]{Ruszczynski1980}
comes from the Lipschitz property of $\nabla U$ from Assumption (b)
and stepsize rule iii) in (\ref{eq:stepsize}) of Theorem \ref{thm:convergence}.\end{IEEEproof}
\begin{lem}
\label{Lemma_Lip_prop}Given problem (\ref{eq:social-formulation})
under Assumptions (a)-(c), suppose that the stepsizes $\{\gamma^{t}\}$
and $\{\rho^{t}\}$ are chosen according to (\ref{eq:stepsize}).
Let $\{\mathbf{x}^{t}\}$ be the sequence generated by Algorithm 1.
Then, there exists a constant $\hat{L}$ such that
\[
\left\Vert \hat{\mathbf{x}}(\mathbf{x}^{t_{1}},\boldsymbol{\xi}^{t_{1}})-\hat{\mathbf{x}}(\mathbf{x}^{t_{2}},\boldsymbol{\xi}^{t_{2}})\right\Vert \leq\hat{L}\left\Vert \mathbf{x}^{t_{1}}-\mathbf{x}^{t_{2}}\right\Vert +e(t_{1},t_{2}),
\]
and $\lim_{t_{1},t_{2}\rightarrow\infty}e(t_{1},t_{2})=0$ w.p.1.\end{lem}
\begin{IEEEproof}
We assume w.l.o.g. that $t_{2}>t_{1}$; for notational simplicity,
we also define $\hat{\mathbf{x}}_{i}^{t}\triangleq\hat{\mathbf{x}}_{i}(\mathbf{x}^{t},\boldsymbol{\xi}^{t})$,
for $t=t_{1}$ and $t=t_{2}.$ It follows from the first-order optimality
condition that \cite{Scutari2012a}\begin{subequations}\label{eq:lip-1}
\begin{align}
\bigl\langle\mathbf{x}_{i}-\hat{\mathbf{x}}_{i}^{t_{1}},\nabla_{i}\hat{f}_{i}(\hat{\mathbf{x}}_{i}^{t_{1}};\mathbf{x}^{t_{1}},\boldsymbol{\xi}^{t_{1}})\bigr\rangle & \geq0,\label{eq:lip-1a}\\
\bigl\langle\mathbf{x}_{i}-\hat{\mathbf{x}}_{i}^{t_{2}},\nabla_{i}\hat{f}_{i}(\hat{\mathbf{x}}_{i}^{t_{2}};\mathbf{x}^{t_{2}},\boldsymbol{\xi}^{t_{2}})\bigr\rangle & \geq0.\label{eq:lip-1b}
\end{align}
\end{subequations}Setting $\mathbf{x}_{i}=\hat{\mathbf{x}}_{i}(\mathbf{x}^{t_{2}},\boldsymbol{\xi}^{t_{2}})$
in (\ref{eq:lip-1a}) and $\mathbf{x}_{i}=\hat{\mathbf{x}}_{i}(\mathbf{x}^{t_{1}},\boldsymbol{\xi}^{t_{1}})$
in (\ref{eq:lip-1b}), and adding the two inequalities, we have
\begin{align}
0 & \geq\bigl\langle\hat{\mathbf{x}}_{i}^{t_{1}}-\hat{\mathbf{x}}_{i}^{t_{2}},\nabla_{i}\hat{f}_{i}(\hat{\mathbf{x}}_{i}^{t_{1}};\mathbf{x}^{t_{1}},\boldsymbol{\xi}^{t_{1}})-\nabla_{i}\hat{f}_{i}(\hat{\mathbf{x}}_{i}^{t_{2}};\mathbf{x}^{t_{2}},\boldsymbol{\xi}^{t_{2}})\bigr\rangle\nonumber \\
 & =\bigl\langle\hat{\mathbf{x}}_{i}^{t_{1}}-\hat{\mathbf{x}}_{i}^{t_{2}},\nabla_{i}\hat{f}_{i}(\hat{\mathbf{x}}_{i}^{t_{1}};\mathbf{x}^{t_{1}},\boldsymbol{\xi}^{t_{1}})-\nabla_{i}\hat{f}_{i}(\hat{\mathbf{x}}_{i}^{t_{1}};\mathbf{x}^{t_{2}},\boldsymbol{\xi}^{t_{2}})\bigr\rangle\nonumber \\
 & +\bigl\langle\hat{\mathbf{x}}_{i}^{t_{1}}-\hat{\mathbf{x}}_{i}^{t_{2}},\nabla_{i}\hat{f}_{i}(\hat{\mathbf{x}}_{i}^{t_{1}};\mathbf{x}^{t_{2}},\boldsymbol{\xi}^{t_{2}})-\nabla_{i}\hat{f}_{i}(\hat{\mathbf{x}}_{i}^{t_{2}};\mathbf{x}^{t_{2}},\boldsymbol{\xi}^{t_{2}})\bigr\rangle.\label{eq:monotone-operator}
\end{align}

The first term in (\ref{eq:monotone-operator}) can be lower bounded
as follows: \begin{subequations}
\begin{align}
 & \bigl\langle\hat{\mathbf{x}}_{i}^{t_{1}}-\hat{\mathbf{x}}_{i}^{t_{2}},\nabla_{i}\hat{f}_{i}(\hat{\mathbf{x}}_{i}^{t_{1}};\mathbf{x}^{t_{1}},\boldsymbol{\xi}^{t_{1}})-\nabla_{i}\hat{f}_{i}(\hat{\mathbf{x}}_{i}^{t_{1}};\mathbf{x}^{t_{2}},\boldsymbol{\xi}^{t_{2}})\bigr\rangle\nonumber \\
=\; & \rho^{t_{1}}{\textstyle \sum_{j\in\mathcal{C}_{i}^{t_{1}}}}\bigl\langle\hat{\mathbf{x}}_{i}^{t_{1}}-\hat{\mathbf{x}}_{i}^{t_{2}},\nonumber \\
 & \qquad\qquad\nabla_{i}f_{j}(\hat{\mathbf{x}}_{i}^{t_{1}},\mathbf{x}_{-i}^{t_{1}},\boldsymbol{\xi}^{t_{1}})-\nabla_{i}f_{j}(\mathbf{x}_{i}^{t_{1}},\mathbf{x}_{-i}^{t_{1}},\boldsymbol{\xi}^{t_{1}})\bigr\rangle\nonumber \\
 & -\rho^{t_{2}}{\textstyle \sum_{j\in\mathcal{C}_{i}^{t_{2}}}}\bigl\langle\hat{\mathbf{x}}_{i}^{t_{1}}-\hat{\mathbf{x}}_{i}^{t_{2}},\nonumber \\
 & \qquad\qquad\nabla_{i}f_{j}(\hat{\mathbf{x}}_{i}^{t_{1}},\mathbf{x}_{-i}^{t_{2}},\boldsymbol{\xi}^{t_{2}})-\nabla_{i}f_{j}(\mathbf{x}_{i}^{t_{2}},\mathbf{x}_{-i}^{t_{2}},\boldsymbol{\xi}^{t_{2}})\bigr\rangle\nonumber \\
 & +\bigl\langle\hat{\mathbf{x}}_{i}^{t_{1}}-\hat{\mathbf{x}}_{i}^{t_{2}},\mathbf{f}_{i}^{t_{1}}-\mathbf{f}_{i}^{t_{2}}\bigr\rangle-\tau_{i}\bigl\langle\hat{\mathbf{x}}_{i}^{t_{1}}-\hat{\mathbf{x}}_{i}^{t_{2}},\mathbf{x}_{i}^{t_{1}}-\mathbf{x}_{i}^{t_{2}}\bigr\rangle\label{eq:term-b-1}
\end{align}
where in (\ref{eq:term-b-1}) we used (\ref{eq:update-rule-incremental-gradient-1}).
Invoking the Lipschitz continuity of $\nabla f_{j}(\bullet,\mathbf{x}_{-i}^{t},\boldsymbol{\xi}^{t})$,
we can get a lower bound for (\ref{eq:term-b-1}): 
\begin{align}
 & \bigl\langle\hat{\mathbf{x}}_{i}^{t_{1}}-\hat{\mathbf{x}}_{i}^{t_{2}},\nabla_{i}\hat{f}_{i}(\hat{\mathbf{x}}_{i}^{t_{1}};\mathbf{x}^{t_{1}},\boldsymbol{\xi}^{t_{1}})-\nabla_{i}\hat{f}_{i}(\hat{\mathbf{x}}_{i}^{t_{1}};\mathbf{x}^{t_{2}},\boldsymbol{\xi}^{t_{2}})\bigr\rangle\nonumber \\
\geq\; & -\rho^{t_{1}}{\textstyle \sum_{j\in\mathcal{C}_{i}^{t_{1}}}}\bigl\Vert\hat{\mathbf{x}}_{i}^{t_{1}}-\hat{\mathbf{x}}_{i}^{t_{2}}\bigr\Vert\cdot\nonumber \\
 & \qquad\qquad\bigl\Vert\nabla_{i}f_{j}(\hat{\mathbf{x}}_{i}^{t_{1}},\mathbf{x}_{-i}^{t_{1}},\boldsymbol{\xi}^{t_{1}})-\nabla_{i}f_{j}(\mathbf{x}_{i}^{t_{1}},\mathbf{x}_{-i}^{t_{1}},\boldsymbol{\xi}^{t_{1}})\bigr\Vert\nonumber \\
 & -\rho^{t_{2}}{\textstyle \sum_{j\in\mathcal{C}_{i}^{t_{2}}}}\bigl\Vert\hat{\mathbf{x}}_{i}^{t_{1}}-\hat{\mathbf{x}}_{i}^{t_{2}}\bigr\Vert\cdot\nonumber \\
 & \qquad\qquad\bigl\Vert\nabla_{i}f_{j}(\hat{\mathbf{x}}_{i}^{t_{1}},\mathbf{x}_{-i}^{t_{2}},\boldsymbol{\xi}^{t_{2}})-\nabla_{i}f_{j}(\mathbf{x}_{i}^{t_{2}},\mathbf{x}_{-i}^{t_{2}},\boldsymbol{\xi}^{t_{2}})\bigr\Vert\nonumber \\
 & +\bigl\langle\hat{\mathbf{x}}_{i}^{t_{1}}-\hat{\mathbf{x}}_{i}^{t_{2}},\mathbf{f}_{i}^{t_{1}}-\nabla_{i}U(\mathbf{x}^{t_{1}})-\mathbf{f}_{i}^{t_{2}}+\nabla_{i}U(\mathbf{x}^{t_{2}})\bigr\rangle\nonumber \\
 & +\bigl\langle\hat{\mathbf{x}}_{i}^{t_{1}}-\hat{\mathbf{x}}_{i}^{t_{2}},\nabla_{i}U(\mathbf{x}^{t_{1}})-\nabla_{i}U(\mathbf{x}^{t_{2}})\bigr\rangle\nonumber \\
 & -\tau_{i}\bigl\langle\hat{\mathbf{x}}_{i}^{t_{1}}-\hat{\mathbf{x}}_{i}^{t_{2}},\mathbf{x}_{i}^{t_{1}}-\mathbf{x}_{i}^{t_{2}}\bigr\rangle,\label{eq:term-b-2}\\
\geq\; & -\rho^{t_{1}}\Bigl({\textstyle \sum_{j\in\mathcal{I}_{f}}}L_{\nabla f_{j}(\boldsymbol{\xi}^{t_{1}})}\Bigr)\bigl\Vert\hat{\mathbf{x}}_{i}^{t_{1}}-\hat{\mathbf{x}}_{i}^{t_{2}}\bigr\Vert\cdot\bigl\Vert\hat{\mathbf{x}}_{i}^{t_{1}}-\mathbf{x}_{i}^{t_{1}}\bigr\Vert\nonumber \\
 & -\rho^{t_{2}}\Bigl({\textstyle \sum_{j\in\mathcal{I}_{f}}}L_{\nabla f_{j}(\boldsymbol{\xi}^{t_{2}})}\Bigr)\bigl\Vert\hat{\mathbf{x}}_{i}^{t_{1}}-\hat{\mathbf{x}}_{i}^{t_{2}}\bigr\Vert\cdot\bigl\Vert\hat{\mathbf{x}}_{i}^{t_{1}}-\mathbf{x}^{t_{2}}\bigr\Vert\nonumber \\
 & -\bigl\Vert\hat{\mathbf{x}}_{i}^{t_{1}}-\hat{\mathbf{x}}_{i}^{t_{2}}\bigr\Vert(\varepsilon^{t_{1}}+\varepsilon^{t_{2}})\nonumber \\
 & -(L_{\nabla U}+\tau_{\max})\bigl\Vert\hat{\mathbf{x}}_{i}^{t_{1}}-\hat{\mathbf{x}}_{i}^{t_{2}}\bigr\Vert\bigl\Vert\mathbf{x}^{t_{1}}-\mathbf{x}^{t_{2}}\bigr\Vert\label{eq:term-b-3}\\
\geq\; & -\rho^{t_{1}}\Bigl({\textstyle \sum_{j\in\mathcal{I}_{f}}}L_{\nabla f_{j}(\boldsymbol{\xi}^{t_{1}})}\Bigr)C_{x}\bigl\Vert\hat{\mathbf{x}}_{i}^{t_{1}}-\hat{\mathbf{x}}_{i}^{t_{2}}\bigr\Vert\nonumber \\
 & -\rho^{t_{2}}\Bigl({\textstyle \sum_{j\in\mathcal{I}_{f}}}L_{\nabla f_{j}(\boldsymbol{\xi}^{t_{2}})}\Bigr)C_{x}\bigl\Vert\hat{\mathbf{x}}_{i}^{t_{1}}-\hat{\mathbf{x}}_{i}^{t_{2}}\bigr\Vert\nonumber \\
 & -\bigl\Vert\hat{\mathbf{x}}_{i}^{t_{1}}-\hat{\mathbf{x}}_{i}^{t_{2}}\bigr\Vert(\varepsilon^{t_{1}}+\varepsilon^{t_{2}})\nonumber \\
 & -(L_{\nabla U}+\tau_{\max})\bigl\Vert\hat{\mathbf{x}}_{i}^{t_{1}}-\hat{\mathbf{x}}_{i}^{t_{2}}\bigr\Vert\bigl\Vert\mathbf{x}^{t_{1}}-\mathbf{x}^{t_{2}}\bigr\Vert,\label{eq:strong-convexity-2}
\end{align}
\end{subequations}where (\ref{eq:term-b-3}) comes from the Lipschitz
continuity of $\nabla f_{j}(\bullet,\mathbf{x}_{-i}^{t},\boldsymbol{\xi}^{t})$,
with $\varepsilon^{t}\triangleq\left\Vert \mathbf{f}^{t}-\nabla U(\mathbf{x}^{t})\right\Vert $
and $\tau_{\max}=\max_{1\leq i\leq I}\tau_{i}<\infty$, and we used
the boundedness of the constraint set $\mathcal{X}$ ($\bigl\Vert\mathbf{x}-\mathbf{y}\bigr\Vert\leq C_{x}$
for some $C_{x}<\infty$ and all $\mathbf{x},\mathbf{y}\in\mathcal{X}$)
and the Lipschitz continuity of $\nabla U(\bullet)$ in (\ref{eq:strong-convexity-2}). 

The second term in (\ref{eq:monotone-operator}) can be bounded as:
\begin{align}
 & \bigl\langle\hat{\mathbf{x}}_{i}^{t_{1}}-\hat{\mathbf{x}}_{i}^{t_{2}},\nabla_{i}\hat{f}(\hat{\mathbf{x}}_{i}^{t_{1}};\mathbf{x}^{t_{2}},\boldsymbol{\xi}^{t_{2}})-\nabla_{i}\hat{f}(\hat{\mathbf{x}}_{i}^{t_{2}};\mathbf{x}^{t_{2}},\boldsymbol{\xi}^{t_{2}})\bigr\rangle\nonumber \\
=\; & \rho^{t_{2}}\sum_{j\in\mathcal{C}_{i}^{t_{2}}}\bigl\langle\hat{\mathbf{x}}_{i}^{t_{1}}-\hat{\mathbf{x}}_{i}^{t_{2}},\nabla_{i}f_{j}(\hat{\mathbf{x}}_{i}^{t_{1}},\mathbf{x}_{-i}^{t_{2}},\boldsymbol{\xi}^{t_{2}})-\nabla_{i}f_{j}(\hat{\mathbf{x}}_{i}^{t_{2}},\mathbf{x}_{-i}^{t_{2}},\boldsymbol{\xi}^{t_{2}})\bigr\rangle\nonumber \\
 & +\tau_{i}\bigl\Vert\hat{\mathbf{x}}_{i}^{t_{1}}-\hat{\mathbf{x}}_{i}^{t_{2}}\bigr\Vert^{2}\geq\tau_{\min}\bigl\Vert\hat{\mathbf{x}}_{i}^{t_{1}}-\hat{\mathbf{x}}_{i}^{t_{2}}\bigr\Vert^{2},\label{eq:strong-convexity-1}
\end{align}
where the inequality follows from the definition of $\tau_{\min}$
and the (uniformly) convexity of the functions $f_{j}(\bullet,\mathbf{x}_{-i}^{t},\boldsymbol{\xi}^{t})$.

Combining the inequalities (\ref{eq:monotone-operator}), (\ref{eq:strong-convexity-2})
and (\ref{eq:strong-convexity-1}), we have
\[
\begin{split}\bigl\Vert\hat{\mathbf{x}}_{i}^{t_{1}}-\hat{\mathbf{x}}_{i}^{t_{2}}\bigr\Vert\leq\; & (L_{\nabla U}+\tau_{\max})\tau_{\min}^{-1}\bigl\Vert\mathbf{x}^{t_{1}}-\mathbf{x}^{t_{2}}\bigr\Vert\\
 & +\tau_{\min}^{-1}C_{x}\rho^{t_{1}}\Bigl({\textstyle \sum_{j\in\mathcal{I}_{f}}}L_{\nabla f_{j}(\boldsymbol{\xi}^{t_{1}})}\Bigr)\\
 & +\tau_{\min}^{-1}C_{x}\rho^{t_{2}}\Bigl({\textstyle \sum_{j\in\mathcal{I}_{f}}}L_{\nabla f_{j}(\boldsymbol{\xi}^{t_{2}})}\Bigr)\\
 & +\tau_{\min}^{-1}(\varepsilon^{t_{1}}+\varepsilon^{t_{2}}),
\end{split}
\]
which leads to the desired (asymptotic) Lipschitz property:
\[
\bigl\Vert\hat{\mathbf{x}}^{t_{1}}-\hat{\mathbf{x}}^{t_{2}}\bigr\Vert\leq\hat{L}\bigl\Vert\mathbf{x}^{t_{1}}-\mathbf{x}^{t_{2}}\bigr\Vert+e(t_{1},t_{2}),
\]
with $\hat{L}\triangleq I\,\tau_{\min}^{-1}(L_{\nabla U}+\tau_{\max})$
and
\[
\begin{array}{l}
e(t_{1},t_{2})\triangleq I\,\tau_{\min}^{-1}\Bigl((\varepsilon^{t_{1}}+\varepsilon^{t_{2}})+\smallskip\\
\quad\quad+C_{x}\bigl(\rho^{t_{1}}\sum_{j\in\mathcal{I}_{f}}L_{\nabla f_{j}(\boldsymbol{\xi}^{t_{1}})}+\rho^{t_{2}}\sum_{j\in\mathcal{I}_{f}}L_{\nabla f_{j}(\boldsymbol{\xi}^{t_{2}})}\bigr)\Bigr).
\end{array}
\]
In view of Lemma \ref{lem:incremental-gradient} and (\ref{eq:stepsize-4}),
it is easy to check that $\lim_{t_{1}\rightarrow\infty,t_{2}\rightarrow\infty}e(t_{1},t_{2})=0$
w.p.1.
\end{IEEEproof}
\medskip{}

\noindent\textbf{Proof of Theorem \ref{thm:convergence}. }Invoking
the first-order optimality conditions of (\ref{eq:approximation-problem}),
we have
\[
\begin{split} & \rho^{t}\bigl\langle\mathbf{x}_{i}^{t}-\hat{\mathbf{x}}_{i}^{t},{\textstyle \sum_{j\in\mathcal{C}_{i}^{t}}}\nabla_{i}f_{j}(\hat{\mathbf{x}}_{i}^{t},\mathbf{x}_{-i}^{t},\boldsymbol{\xi}^{t})+\boldsymbol{\pi}_{i}(\mathbf{x}^{t},\boldsymbol{\xi}^{t})\bigr\rangle\\
 & +(1-\rho^{t})\bigl\langle\mathbf{x}_{i}^{t}-\hat{\mathbf{x}}_{i}^{t},\mathbf{f}_{i}^{t-1}\bigr\rangle+\tau_{i}\bigl\langle\mathbf{x}_{i}^{t}-\hat{\mathbf{x}}_{i}^{t},\hat{\mathbf{x}}_{i}^{t}-\mathbf{x}_{i}^{t}\bigr\rangle\\
=\; & \rho^{t}\,{\textstyle \sum_{j\in\mathcal{C}_{i}^{t}}}\bigl\langle\mathbf{x}_{i}^{t}-\hat{\mathbf{x}}_{i}^{t},\nabla_{i}f_{j}(\hat{\mathbf{x}}_{i}^{t},\mathbf{x}_{-i}^{t},\boldsymbol{\xi}^{t})-\nabla_{i}f_{j}(\mathbf{x}_{i}^{t},\mathbf{x}_{-i}^{t},\boldsymbol{\xi}^{t})\bigr\rangle\\
 & +\bigl\langle\mathbf{x}_{i}^{t}-\hat{\mathbf{x}}_{i}^{t},\mathbf{f}_{i}^{t}\bigr\rangle-\tau_{i}\bigl\Vert\hat{\mathbf{x}}_{i}^{t}-\mathbf{x}_{i}^{t}\bigr\Vert^{2}\geq0,
\end{split}
\]
which together with the convexity of $\sum_{j\in\mathcal{C}_{i}^{t}}f_{j}(\bullet,\mathbf{x}_{-i}^{t},\boldsymbol{\xi}^{t})$
leads to 
\begin{equation}
\bigl\langle\hat{\mathbf{x}}_{i}^{t}-\mathbf{x}_{i}^{t},\mathbf{f}_{i}^{t}\bigr\rangle\leq-\tau_{\min}\bigl\Vert\hat{\mathbf{x}}_{i}^{t}-\mathbf{x}_{i}^{t}\bigr\Vert^{2}.\label{eq:upper_bound}
\end{equation}

It follows from the descent lemma on $U$ that
\begin{align}
U(\mathbf{x}^{t+1})\leq\; & U(\mathbf{x}^{t})+\gamma^{t+1}\bigl\langle\hat{\mathbf{x}}^{t}-\mathbf{x}^{t},\nabla U(\mathbf{x}^{t})\bigr\rangle\nonumber \\
 & +L_{\nabla U}(\gamma^{t+1})^{2}\bigl\Vert\hat{\mathbf{x}}^{t}-\mathbf{x}^{t}\bigr\Vert^{2}\nonumber \\
=\; & U(\mathbf{x}^{t})+\gamma^{t+1}\bigl\langle\hat{\mathbf{x}}^{t}-\mathbf{x}^{t},\nabla U(\mathbf{x}^{t})-\mathbf{f}^{t}+\mathbf{f}^{t}\bigr\rangle\nonumber \\
 & +L_{\nabla U}(\gamma^{t+1})^{2}\bigl\Vert\hat{\mathbf{x}}^{t}-\mathbf{x}^{t}\bigr\Vert^{2}\nonumber \\
\leq\; & U(\mathbf{x}^{t})-\gamma^{t+1}(\tau_{\min}-L_{\nabla U}\gamma^{t+1})\bigl\Vert\hat{\mathbf{x}}^{t}-\mathbf{x}^{t}\bigr\Vert^{2}\nonumber \\
 & +\gamma^{t+1}\bigl\Vert\hat{\mathbf{x}}^{t}-\mathbf{x}^{t}\bigr\Vert\bigl\Vert\nabla U(\mathbf{x}^{t})-\mathbf{f}^{t}\bigr\Vert,\label{eq:descent_lemma}
\end{align}
where in the last inequality we used (\ref{eq:upper_bound}). Let
us show by contradiction that $\lim\inf_{t\rightarrow\infty}\bigl\Vert\hat{\mathbf{x}}^{t}-\mathbf{x}^{t}\bigr\Vert=0$
w.p.1. Suppose $\lim\inf_{t\rightarrow\infty}\bigl\Vert\hat{\mathbf{x}}^{t}-\mathbf{x}^{t}\bigr\Vert\geq\chi>0$
with a positive probability. Then we can find a realization such that
at the same time $\bigl\Vert\hat{\mathbf{x}}^{t}-\mathbf{x}^{t}\bigr\Vert\geq\chi>0$
for all $t$ and $\lim_{t\rightarrow\infty}\bigl\Vert\nabla U(\mathbf{x}^{t})-\mathbf{f}^{t}\bigr\Vert=0$;
we focus next on such a realization. Using $\bigl\Vert\hat{\mathbf{x}}^{t}-\mathbf{x}^{t}\bigr\Vert\geq\chi>0$,
the inequality (\ref{eq:descent_lemma}) is equivalent to
\begin{equation}
\begin{array}{l}
U(\mathbf{x}^{t+1})-U(\mathbf{x}^{t})\leq\smallskip\\
\;-\gamma^{t+1}\left(\tau_{\min}-L_{\nabla U}\gamma^{t+1}-\frac{1}{\chi}\left\Vert \nabla U(\mathbf{x}^{t})-\mathbf{f}^{t}\right\Vert \right)\bigl\Vert\hat{\mathbf{x}}^{t}-\mathbf{x}^{t}\bigr\Vert^{2}.
\end{array}\label{eq:proof-descent}
\end{equation}
Since $\lim_{t\rightarrow\infty}\bigl\Vert\nabla U(\mathbf{x}^{t})-\mathbf{f}^{t}\bigr\Vert=0$,
there exists a $t_{0}$ sufficiently large such that
\begin{equation}
\tau_{\min}-L_{\nabla U}\gamma^{t+1}-\frac{1}{\chi}\left\Vert \nabla U(\mathbf{x}^{t})-\mathbf{f}^{t}\right\Vert \geq\bar{\tau}>0,\quad\forall\, t\geq t_{0}.\label{eq:proof-t0}
\end{equation}
Therefore, it follows from (\ref{eq:proof-descent}) and (\ref{eq:proof-t0})
that
\begin{equation}
U(\mathbf{x}^{t})-U(\mathbf{x}^{t_{0}})\leq-\bar{\tau}\chi^{2}{\textstyle \sum_{n=t_{0}}^{t}}\gamma^{n+1},\label{eq:proof-counter-bound}
\end{equation}
which, in view of $\sum_{n=t_{0}}^{\infty}\gamma^{n+1}=\infty$, contradicts
the boundedness of $\{U(\mathbf{x}^{t})\}$. Therefore it must be
$\lim\inf_{t\rightarrow\infty}\left\Vert \hat{\mathbf{x}}^{t}-\mathbf{x}^{t}\right\Vert =0$
w.p.1.

We prove now that $\lim\sup_{t\rightarrow\infty}\left\Vert \hat{\mathbf{x}}^{t}-\mathbf{x}^{t}\right\Vert =0$
w.p.1. Assume $\lim\sup_{t\rightarrow\infty}\left\Vert \hat{\mathbf{x}}^{t}-\mathbf{x}^{t}\right\Vert >0$
with some positive probability. We focus next on a realization along
with $\lim\sup_{t\rightarrow\infty}\left\Vert \hat{\mathbf{x}}^{t}-\mathbf{x}^{t}\right\Vert >0$,
 $\lim_{t\rightarrow\infty}\bigl\Vert\nabla U(\mathbf{x}^{t})-\mathbf{f}^{t}\bigr\Vert=0$,
$\lim\inf_{t\rightarrow\infty}\bigl\Vert\hat{\mathbf{x}}^{t}-\mathbf{x}^{t}\bigr\Vert=0$,
and $\lim_{t_{i},t_{2}\rightarrow\infty}e(t_{1},t_{2})=0$, where
$e(t_{1},t_{2})$ is defined in Lemma \ref{Lemma_Lip_prop}. It follows
from $\lim\sup_{t\rightarrow\infty}\left\Vert \hat{\mathbf{x}}^{t}-\mathbf{x}^{t}\right\Vert >0$
and $\lim\inf_{t\rightarrow\infty}\bigl\Vert\hat{\mathbf{x}}^{t}-\mathbf{x}^{t}\bigr\Vert=0$
that there exists a $\delta>0$ such that $\left\Vert \triangle\mathbf{x}^{t}\right\Vert \geq2\delta$
(with $\triangle\mathbf{x}^{t}\triangleq\hat{\mathbf{x}}^{t}-\mathbf{x}^{t}$)
for infinitely many $t$ and also $\left\Vert \triangle\mathbf{x}^{t}\right\Vert <\delta$
for infinitely many $t$. Therefore, one can always find an infinite
set of indexes, say $\mathcal{T}$, having the following properties:
for any $t\in\mathcal{T}$, there exists an integer $i_{t}>t$ such
that
\begin{equation}
\begin{array}{l}
\left\Vert \triangle\mathbf{x}^{t}\right\Vert <\delta,\quad\left\Vert \triangle\mathbf{x}^{i_{t}}\right\Vert >2\delta,\smallskip\\
\delta\leq\left\Vert \triangle\mathbf{x}^{n}\right\Vert \leq2\delta,\quad t<n<i_{t}.
\end{array}\label{eq:proof-set-T}
\end{equation}
Given the above bounds, the following holds: for all $t\in\mathcal{T}$,
\begin{align}
\delta & \leq\left\Vert \triangle\mathbf{x}^{i_{t}}\right\Vert -\left\Vert \triangle\mathbf{x}^{t}\right\Vert \nonumber \\
 & \leq\left\Vert \triangle\mathbf{x}^{i_{t}}-\triangle\mathbf{x}^{t}\right\Vert =\left\Vert (\hat{\mathbf{x}}^{i_{t}}-\mathbf{x}^{i_{t}})-(\hat{\mathbf{x}}^{t}-\mathbf{x}^{t})\right\Vert \nonumber \\
 & \leq\bigl\Vert\hat{\mathbf{x}}^{i_{t}}-\hat{\mathbf{x}}^{t}\bigr\Vert+\bigl\Vert\mathbf{x}^{i_{t}}-\mathbf{x}^{t}\bigr\Vert\nonumber \\
 & \leq(1+\hat{L})\bigl\Vert\mathbf{x}^{i_{t}}-\mathbf{x}^{t}\bigr\Vert+e(i_{t},t)\nonumber \\
 & \leq(1+\hat{L}){\textstyle \sum_{n=t}^{i_{t}-1}}\gamma^{n+1}\left\Vert \triangle\mathbf{x}^{n}\right\Vert +e(i_{t},t)\nonumber \\
 & \leq2\delta(1+\hat{L}){\textstyle \sum_{n=t}^{i_{t}-1}}\gamma^{n+1}+e(i_{t},t),\label{eq:proof-liminf}
\end{align}
implying that\vspace{-0.2cm} 
\begin{equation}
\underset{\mathcal{T}\ni t\rightarrow\infty}{\lim\inf}\;{\textstyle \sum_{n=t}^{i_{t}-1}}\gamma^{n+1}\geq\bar{\delta}_{1}\triangleq\frac{1}{2(1+\hat{L})}>0.\label{eq:proof-stepsize-lowerbound}
\end{equation}

Proceeding as in (\ref{eq:proof-liminf}), we also have: for all $t\in\mathcal{T}$,
\[
\begin{split}\left\Vert \triangle\mathbf{x}^{t+1}\right\Vert -\left\Vert \triangle\mathbf{x}^{t}\right\Vert  & \leq\left\Vert \triangle\mathbf{x}^{t+1}-\triangle\mathbf{x}^{t}\right\Vert \\
 & \leq(1+\hat{L})\gamma^{t+1}\left\Vert \triangle\mathbf{x}^{t}\right\Vert +e(t,t+1),
\end{split}
\]
which leads to
\begin{equation}
(1+(1+\hat{L})\gamma^{t+1})\left\Vert \triangle\mathbf{x}^{t}\right\Vert +e(t,t+1)\geq\left\Vert \triangle\mathbf{x}^{t+1}\right\Vert \geq\delta,\label{eq:lower_bound_delta}
\end{equation}
where the second inequality follows from (\ref{eq:proof-set-T}).
It follows from (\ref{eq:lower_bound_delta}) that there exists a
$\bar{\delta}_{2}>0$ such that for sufficiently large $t\in\mathcal{T}$,
\begin{equation}
\left\Vert \triangle\mathbf{x}^{t}\right\Vert \geq\frac{\delta-e(t,t+1)}{1+(1+\hat{L})\gamma^{t+1}}\geq\bar{\delta}_{2}>0.\label{eq:proof-error-lower-bound}
\end{equation}
Here after we assume w.l.o.g. that (\ref{eq:proof-error-lower-bound})
holds for all $t\in\mathcal{T}$ (in fact one can always restrict
$\{\mathbf{x}^{t}\}_{t\in\mathcal{T}}$ to a proper subsequence). 

We show now that (\ref{eq:proof-stepsize-lowerbound}) is in contradiction
with the convergence of $\{U(\mathbf{x}^{t})\}$. Invoking (\ref{eq:descent_lemma}),
we have: for all $t\in\mathcal{T}$,\vspace{-0.2cm}
\begin{align}
 & U(\mathbf{x}^{t+1})-U(\mathbf{x}^{t})\nonumber \\
 & \quad\leq-\gamma^{t+1}\left(\tau_{\min}-L_{\nabla U}\gamma^{t+1}\right)\bigl\Vert\hat{\mathbf{x}}^{t}-\mathbf{x}^{t}\bigr\Vert^{2}\nonumber \\
 & \quad\qquad+\gamma^{t+1}\delta\bigl\Vert\nabla U(\mathbf{x}^{t})-\mathbf{f}^{t}\bigr\Vert\nonumber \\
 & \quad\leq-\gamma^{t+1}\left(\tau_{\min}-L_{\nabla U}\gamma^{t+1}-\frac{\bigl\Vert\nabla U(\mathbf{x}^{t})-\mathbf{f}^{t}\bigr\Vert}{\delta}\right)\nonumber \\
 & \quad\qquad\cdot\bigl\Vert\hat{\mathbf{x}}^{t}-\mathbf{x}^{t}\bigr\Vert^{2}+\gamma^{t+1}\delta\bigl\Vert\nabla U(\mathbf{x}^{t})-\mathbf{f}^{t}\bigr\Vert^{2},\label{eq:proof-descent-lemman-t}
\end{align}
and for $t<n<i_{t}$,
\begin{align}
 & U(\mathbf{x}^{n+1})-U(\mathbf{x}^{n})\nonumber \\
 & \quad\leq-\gamma^{n+1}\left(\tau_{\min}-L_{\nabla U}\gamma^{n+1}-\frac{\bigl\Vert\nabla U(\mathbf{x}^{n})-\mathbf{f}^{n}\bigr\Vert}{\bigl\Vert\hat{\mathbf{x}}^{n}-\mathbf{x}^{n}\bigr\Vert}\right)\nonumber \\
 & \quad\qquad\cdot\bigl\Vert\hat{\mathbf{x}}^{n}-\mathbf{x}^{n}\bigr\Vert^{2}\nonumber \\
 & \quad\leq-\gamma^{n+1}\left(\tau_{\min}-L_{\nabla U}\gamma^{n+1}-\frac{\bigl\Vert\nabla U(\mathbf{x}^{n})-\mathbf{f}^{n}\bigr\Vert}{\delta}\right)\nonumber \\
 & \qquad\cdot\bigl\Vert\hat{\mathbf{x}}^{n}-\mathbf{x}^{n}\bigr\Vert^{2},\label{eq:proof-descent-lemma-bigger-t}
\end{align}
where the last inequality follows from  (\ref{eq:proof-set-T}). Adding
(\ref{eq:proof-descent-lemman-t}) and (\ref{eq:proof-descent-lemma-bigger-t})
over $n=t+1,\ldots,i_{t}-1$ and, for $t\in\mathcal{T}$ sufficiently
large (so that $\tau_{\min}-L_{\nabla U}\gamma^{t+1}-\delta^{-1}\bigl\Vert\nabla U(\mathbf{x}^{n})-\mathbf{f}^{n}\bigr\Vert\geq\hat{\tau}>0$
and $\bigl\Vert\nabla U(\mathbf{x}^{t})-\mathbf{f}^{t}\bigr\Vert<\hat{\tau}\bar{\delta}_{2}^{2}/\delta$),
we have
\begin{align}
 & U(\mathbf{x}^{i_{t}})-U(\mathbf{x}^{t})\nonumber \\
 & \quad\overset{(a)}{\leq}\;-\hat{\tau}\,{\textstyle \sum_{n=t}^{i_{t}-1}}\gamma^{n+1}\bigl\Vert\hat{\mathbf{x}}^{n}-\mathbf{x}^{n}\bigr\Vert^{2}+\gamma^{t+1}\delta\,\bigl\Vert\nabla U(\mathbf{x}^{t})-\mathbf{f}^{t}\bigr\Vert\nonumber \\
 & \quad\overset{(b)}{\leq}\;-\hat{\tau}\,\bar{\delta}_{2}^{2}\,{\textstyle \sum_{n=t+1}^{i_{t}-1}}\gamma^{n+1}-\gamma^{t+1}\left(\hat{\tau}\bar{\delta}_{2}^{2}-\delta\,\bigl\Vert\nabla U(\mathbf{x}^{t})-\mathbf{f}^{t}\bigr\Vert\right)\nonumber \\
 & \quad\overset{(c)}{\leq}\;-\hat{\tau}\,\bar{\delta}_{2}^{2}\,{\textstyle \sum_{n=t+1}^{i_{t}-1}}\gamma^{n+1},\label{eq:proof-convergence-contradiction}
\end{align}
where (a) follows from $\tau_{\min}-L_{\nabla U}\gamma^{t+1}-\delta^{-1}\bigl\Vert\nabla U(\mathbf{x}^{n})-\mathbf{f}^{n}\bigr\Vert\geq\hat{\tau}>0$;
(b) is due to (\ref{eq:proof-error-lower-bound}); and in (c) we used
$\bigl\Vert\nabla U(\mathbf{x}^{t})-\mathbf{f}^{t}\bigr\Vert<\hat{\tau}\bar{\delta}_{2}^{2}/\delta$.
Since $\{U(\mathbf{x}^{t})\}$ converges, it must be $\underset{\mathcal{T}\ni t\rightarrow\infty}{\lim\inf}\;{\textstyle \sum_{n=t+1}^{i_{t}-1}}\gamma^{n+1}=0$,
which contradicts (\ref{eq:proof-stepsize-lowerbound}). Therefore,
it must be $\lim\sup_{t\rightarrow\infty}\bigl\Vert\hat{\mathbf{x}}^{t}-\mathbf{x}^{t}\bigr\Vert=0$
w.p.1.

Finally, let us prove that every limit point of the sequence $\left\{ \mathbf{x}^{t}\right\} $
is a stationary solution of (\ref{eq:social-formulation}). Let $\mathbf{x}^{\infty}$
be the limit point of the convergent subsequence $\left\{ \mathbf{x}^{t}\right\} _{t\in\mathcal{T}}$.
Taking the limit of (\ref{eq:lip-1}) over the index set $\mathcal{T}$,
we have
\begin{align}
 & \lim_{\mathcal{T}\ni t\rightarrow\infty}\left<\mathbf{x}_{i}-\hat{\mathbf{x}}_{i}^{t},\nabla_{i}\hat{f}_{i}\left(\hat{\mathbf{x}}_{i}^{t};\mathbf{x}^{t},\boldsymbol{\xi}^{t}\right)\right>\nonumber \\
 & =\lim_{\mathcal{T}\ni t\rightarrow\infty}\left\langle \mathbf{x}_{i}-\hat{\mathbf{x}}_{i}^{t},\right.\nonumber \\
 & \quad\,\,\mathbf{f}_{i}^{t}+\tau_{i}\left(\hat{\mathbf{x}}_{i}^{t}-\mathbf{x}_{i}^{t}\right)\nonumber \\
 & \quad\left.+\rho^{t}{\textstyle \sum_{j\in\mathcal{C}_{i}^{t}}}\left(\nabla_{i}f_{j}(\hat{\mathbf{x}}_{i}^{t},\mathbf{x}_{-i}^{t},\boldsymbol{\xi}^{t})-\nabla_{i}f_{j}(\mathbf{x}_{i}^{t},\mathbf{x}_{-i}^{t},\boldsymbol{\xi}^{t})\right)\right\rangle \nonumber \\
 & \quad=\bigl\langle\mathbf{x}_{i}-\mathbf{x}_{i}^{\infty},\nabla U(\mathbf{x}_{i}^{\infty})\bigr\rangle\geq0,\;\forall\,\mathbf{x}_{i}\in\mathcal{X}_{i},\label{eq:limit-4}
\end{align}
where the last equality follows from: i) $\lim_{t\rightarrow\infty}\bigl\Vert\nabla U(\mathbf{x}^{t})-\mathbf{f}^{t}\bigr\Vert=0$
(cf. Lemma \ref{lem:incremental-gradient}); ii) $\lim{}_{t\rightarrow\infty}\bigl\Vert\hat{\mathbf{x}}_{i}^{t}-\mathbf{x}^{t}\bigr\Vert=0$;
and iii) the following 
\begin{align}
 & \bigl\Vert\rho^{t}{\textstyle \sum_{j\in\mathcal{C}_{i}^{t}}}(\nabla_{i}f_{j}(\hat{\mathbf{x}}_{i}^{t},\mathbf{x}_{-i}^{t},\boldsymbol{\xi}^{t})-\nabla_{i}f_{j}(\mathbf{x}_{i}^{t},\mathbf{x}_{-i}^{t},\boldsymbol{\xi}^{t}))\bigr\Vert\nonumber \\
 & \leq C_{x}\rho^{t}{\textstyle \sum_{j\in\mathcal{I}_{f}}}L_{\nabla f_{j}(\boldsymbol{\xi}^{t})}\underset{t\rightarrow\infty}{\longrightarrow}0,\label{eq:diminishes}
\end{align}
where (\ref{eq:diminishes}) follows from the Lipschitz continuity
of $\nabla f_{j}(\mathbf{x},\boldsymbol{\xi})$, the fact $\left\Vert \hat{\mathbf{x}}_{i}^{t}-\mathbf{x}_{i}^{t}\right\Vert \leq C_{x}$,
and (\ref{eq:stepsize-4}). 

Adding (\ref{eq:limit-4}) over $i=1,\ldots,I$, we get the desired
first-order optimality condition: $\bigl\langle\mathbf{x}-\mathbf{x}^{\infty},\nabla U(\mathbf{x}^{\infty})\bigr\rangle\geq0,$
for all $\mathbf{x}\in\mathcal{X}.$ Therefore $\mathbf{x}^{\infty}$
is a stationary point of (\ref{eq:social-formulation}).\qed

\subsection{\label{sec:Proof-of-waterfilling-solution}Proof of Lemma \ref{lem:waterfilling-solution}}

We prove only (\ref{eq:Y_closed_form}). Since (\ref{eq:logdet-regularized-s4-2})
is a convex optimization problem and $\mathcal{Q}$ has a nonempty
interior, strong duality holds for (\ref{eq:logdet-regularized-s4-2})
\cite{boyd2004convex}. The dual function of (\ref{eq:logdet-regularized-s4-2})
is
\begin{equation}
d(\mu)=\max_{\mathbf{Y}\succeq\mathbf{0}}\{\rho\log\det(\mathbf{I}+\mathbf{Y}\mathbf{D}_{1})+\bigl\langle\mathbf{Y},\mathbf{Z}-\mu\mathbf{I}\bigr\rangle\}+\mu P,\label{eq:MIMO-BR-Dual}
\end{equation}
where $\mu\in\{\mu:\mu\succeq0,d(\mu)<+\infty\}$. Denote by $\mathbf{Y}^{\star}(\mu)$
the optimal solution of the maximization problem in (\ref{eq:MIMO-BR-Dual}),
for any given feasible $\mu$. It is easy to see that $d(\mu)=+\infty$
if $\mathbf{Z}-\mu\mathbf{I}\succeq\mathbf{0}$, so $\mu$ is feasible
if and only if $\mathbf{Z}-\mu\mathbf{I}\prec\mathbf{0}$, i.e., 
\[
\mu\begin{cases}
\geq\underline{\mu}=[\lambda_{\max}(\mathbf{Z})]^{+}=0, & \textrm{if }\mathbf{Z}\prec\mathbf{0},\\
>\underline{\mu}=[\lambda_{\max}(\mathbf{Z})]^{+}, & \textrm{otherwise},
\end{cases}
\]
and $\mathbf{Y}^{\star}(\mu)$ is \cite[Prop. 1]{Kim2011}
\[
\mathbf{Y}^{\star}(\mu)=\mathbf{V}(\mu)[\rho\mathbf{I}-\mathbf{D}(\mu)^{-1}]^{+}\mathbf{V}(\mu)^{H},
\]
where $(\mathbf{V}(\mu),\mathbf{\boldsymbol{{\Sigma}}}(\mu))$ is
the generalized eigenvalue decomposition of $(\mathbf{D}_{1},-\mathbf{Z}+\mu\mathbf{I})$.
Invoking \cite[Cor. 28.1.1]{Rockafellar70}, the uniqueness of $\mathbf{Y(Z)}$
comes from the uniqueness of $\mathbf{Y}^{\star}(\mu)$ that was proved
in \cite{Yu2004}.

Now we prove that $\mu^{\star}\leq\overline{\mu}$. First, note that
$d(\mu)\geq\mu P$. Based on the eigenvalue decomposition $\mathbf{Z}=\mathbf{V}_{\mathbf{Z}}\boldsymbol{\Sigma}_{\mathbf{Z}}\mathbf{V}_{\mathbf{Z}}^{H}$,
the following inequalities hold:
\[
\begin{split}\textrm{tr}((\mathbf{Z}-\mu\mathbf{I})^{H}\mathbf{X}) & =\textrm{tr}(\mathbf{V}_{\mathbf{Z}}(\boldsymbol{\Sigma}_{\mathbf{Z}}-\mu\mathbf{I})\mathbf{V}_{\mathbf{Z}}^{H}\mathbf{X})\\
 & \leq(\lambda_{\max}(\boldsymbol{\Sigma}_{\mathbf{Z}})-\mu)\textrm{tr}(\mathbf{X}),
\end{split}
\]
where $\lambda_{\max}(\boldsymbol{\Sigma}_{\mathbf{Z}})=\lambda_{\max}(\mathbf{Z})$.
In other words, $d(\mu)$ is upper bounded by the optimal value of
the following problem:
\begin{equation}
\begin{split}\underset{\mathbf{Y}\succeq\mathbf{0}}{\max}\quad & \rho\log\det(\mathbf{I}+\mathbf{Y}\mathbf{D}_{1})+(\lambda_{\max}(\mathbf{Z})-\mu)\textrm{tr}(\mathbf{Y})+\mu P.\end{split}
\label{eq:MIMO-BR-UB}
\end{equation}

When $\mu\geq\overline{\mu}$, it is not difficult to verify that
the optimal variable of (\ref{eq:MIMO-BR-UB}) is $\mathbf{0}$, and
thus $\mathbf{Y}^{\star}(\mu)=\mathbf{0}$. We show $\mu^{\star}\leq\bar{\mu}$
by discussing two complementary cases: $\bar{\mu}=0$ and $\bar{\mu}>0$.

If $\bar{\mu}=0$, $d(\bar{\mu})=d(0)=\mu P=0$. Since $\mathbf{Y}^{\star}(0)=\mathbf{0}$
and the primal value is also 0, there is no duality gap. From the
definition of saddle point \cite[Sec. 5.4]{boyd2004convex}, $\bar{\mu}=0$
is a dual optimal variable.

If $\bar{\mu}>0$, $d(\mu)\geq\mu P>0$. Assume $\mu^{\star}>\overline{\mu}$.
Then $\mathbf{Y}^{\star}(\mu^{\star})=\mathbf{0}$ is the optimal
variable in (\ref{eq:logdet-regularized-s4-2}) and the optimal value
of (\ref{eq:logdet-regularized-s4-2}) is 0, but this would lead to
a non-zero duality gap and thus contradict the optimality of $\mu^{\star}$.
Therefore $\mu^{\star}\leq\overline{\mu}$. 

\bibliographystyle{bib/MyIEEE}
\bibliography{bib/book_collection,bib/stochastic,bib/unpublished_papers}

\begin{thebibliography}{10}
\providecommand{\url}[1]{#1}
\csname url@samestyle\endcsname
\providecommand{\newblock}{\relax}
\providecommand{\bibinfo}[2]{#2}
\providecommand{\BIBentrySTDinterwordspacing}{\spaceskip=0pt\relax}
\providecommand{\BIBentryALTinterwordstretchfactor}{4}
\providecommand{\BIBentryALTinterwordspacing}{\spaceskip=\fontdimen2\font plus
\BIBentryALTinterwordstretchfactor\fontdimen3\font minus
  \fontdimen4\font\relax}
\providecommand{\BIBforeignlanguage}[2]{{%
\expandafter\ifx\csname l@#1\endcsname\relax
\typeout{** WARNING: IEEEtran.bst: No hyphenation pattern has been}%
\typeout{** loaded for the language `#1'. Using the pattern for}%
\typeout{** the default language instead.}%
\else
\language=\csname l@#1\endcsname
\fi
#2}}
\providecommand{\BIBdecl}{\relax}
\BIBdecl

\bibitem{Yang2013}
\BIBentryALTinterwordspacing
Y.~Yang, G.~Scutari, and D.~P. Palomar, ``{Parallel stochastic decomposition
  algorithms for multi-agent systems},'' \emph{2013 IEEE 14th Workshop on
  Signal Processing Advances in Wireless Communications (SPAWC)}, pp. 180--184,
  Jun. 2013.
\BIBentrySTDinterwordspacing

\bibitem{Robbins1951}
\BIBentryALTinterwordspacing
H.~Robbins and S.~Monro, ``{A Stochastic Approximation Method},'' \emph{The
  Annals of Mathematical Statistics}, vol.~22, no.~3, pp. 400--407, Sep. 1951.
\BIBentrySTDinterwordspacing

\bibitem{KushnerYin2003}
H.~J. Kushner and G.~Yin, \emph{{Stochastic approximation and recursive
  algorithms and applications}}, 2nd~ed.\hskip 1em plus 0.5em minus 0.4em\relax
  Springer-Verlag, 2003, vol.~35.

\bibitem{polyak1987introduction}
B.~Polyak, \emph{{Introduction to optimization}}.\hskip 1em plus 0.5em minus
  0.4em\relax Optimization Software, 1987.

\bibitem{Bertsekas2000}
\BIBentryALTinterwordspacing
D.~P. Bertsekas and J.~N. Tsitsiklis, ``{Gradient convergence in gradient
  methods with errors},'' \emph{SIAM Journal on Optimization}, vol.~10, no.~3,
  pp. 627--642, 2000.
\BIBentrySTDinterwordspacing

\bibitem{Tsitsiklis1986}
\BIBentryALTinterwordspacing
J.~N. Tsitsiklis, D.~P. Bertsekas, and M.~Athans, ``{Distributed asynchronous
  deterministic and stochastic gradient optimization algorithms},'' \emph{IEEE
  Transactions on Automatic Control}, vol.~31, no.~9, pp. 803--812, Sep. 1986.
\BIBentrySTDinterwordspacing

\bibitem{Ermol'ev1972}
\BIBentryALTinterwordspacing
Y.~Ermoliev, ``{On the method of generalized stochastic gradients and
  quasi-Fejer sequences},'' \emph{Cybernetics}, vol.~5, no.~2, pp. 208--220,
  1972.
\BIBentrySTDinterwordspacing

\bibitem{Yousefian2012}
\BIBentryALTinterwordspacing
F.~Yousefian, A.~Nedi\'{c}, and U.~V. Shanbhag, ``{On stochastic gradient and
  subgradient methods with adaptive steplength sequences},'' \emph{Automatica},
  vol.~48, no.~1, pp. 56--67, Jan. 2012.
\BIBentrySTDinterwordspacing

\bibitem{Ermol'ev1977}
\BIBentryALTinterwordspacing
Y.~Ermoliev and P.~I. Verchenko, ``{A linearization method in limiting extremal
  problems},'' \emph{Cybernetics}, vol.~12, no.~2, pp. 240--245, 1977.
\BIBentrySTDinterwordspacing

\bibitem{Ruszczynski1980}
\BIBentryALTinterwordspacing
A.~Ruszczy\'{n}ski, ``{Feasible direction methods for stochastic programming
  problems},'' \emph{Mathematical Programming}, vol.~19, no.~1, pp. 220--229,
  Dec. 1980.
\BIBentrySTDinterwordspacing

\bibitem{Ruszczynski2008}
\BIBentryALTinterwordspacing
------, ``{A merit function approach to the subgradient method with
  averaging},'' \emph{Optimization Methods and Software}, vol.~23, no.~1, pp.
  161--172, Feb. 2008.
\BIBentrySTDinterwordspacing

\bibitem{Nemirovski2009}
\BIBentryALTinterwordspacing
A.~Nemirovski, A.~Juditsky, G.~Lan, and A.~Shapiro, ``{Robust Stochastic
  Approximation Approach to Stochastic Programming},'' \emph{SIAM Journal on
  Optimization}, vol.~19, no.~4, pp. 1574--1609, Jan. 2009.
\BIBentrySTDinterwordspacing

\bibitem{Gupal1974a}
\BIBentryALTinterwordspacing
A.~M. Gupal and L.~G. Bazhenov, ``{Stochastic analog of the conjugant-gradient
  method},'' \emph{Cybernetics}, vol.~8, no.~1, pp. 138--140, 1972.
\BIBentrySTDinterwordspacing

\bibitem{Polyak1992}
\BIBentryALTinterwordspacing
B.~T. Polyak and A.~B. Juditsky, ``{Acceleration of Stochastic Approximation by
  Averaging},'' \emph{SIAM Journal on Control and Optimization}, vol.~30,
  no.~4, pp. 838--855, Jul. 1992.
\BIBentrySTDinterwordspacing

\bibitem{Scutarib}
\BIBentryALTinterwordspacing
G.~Scutari, F.~Facchinei, P.~Song, D.~P. Palomar, and J.-S. Pang,
  ``{Decomposition by Partial Linearization: Parallel Optimization of
  Multi-Agent Systems},'' \emph{IEEE Transactions on Signal Processing},
  vol.~62, no.~3, pp. 641--656, Feb. 2014.
\BIBentrySTDinterwordspacing

\bibitem{Scutari_BigData}
\BIBentryALTinterwordspacing
F.~Facchinei, G.~Scutari, and S.~Sagratella, ``{Parallel Algorithms for Big
  Data Optimization},'' Dec. 2013, submitted to \emph{IEEE Transactions on
  Signal Processing}. [Online]. Available: \url{http://arxiv.org/abs/1402.5521}
\BIBentrySTDinterwordspacing

\bibitem{Scutari_hybrid}
\BIBentryALTinterwordspacing
A.~Daneshmand, F.~Facchinei, V.~Kungurtsev, and G.~Scutari, ``{Hybrid
  Random/Deterministic Parallel Algorithms for Nonconvex Big Data
  Optimization},'' Jun. 2014, submitted to \emph{IEEE Transactions on Signal
  Processing}. [Online]. Available: \url{http://arxiv.org/abs/1407.4504}
\BIBentrySTDinterwordspacing

\bibitem{Razaviyayn_2013}
M.~Razaviyayn, M.~Sanjabi, and Z.-Q. Luo, ``{A stochastic successive
  minimization method for nonsmooth nonconvex optimization},'' Jun. 2013,
  submitted to \emph{Mathematical Programming}.

\bibitem{Robinson1996}
\BIBentryALTinterwordspacing
S.~M. Robinson, ``{Analysis of Sample-Path Optimization},'' \emph{Mathematics
  of Operations Research}, vol.~21, no.~3, pp. 513--528, Aug. 1996.
\BIBentrySTDinterwordspacing

\bibitem{Linderoth2006}
\BIBentryALTinterwordspacing
J.~Linderoth, A.~Shapiro, and S.~Wright, ``{The empirical behavior of sampling
  methods for stochastic programming},'' \emph{Annals of Operations Research},
  vol. 142, no.~1, pp. 215--241, Feb. 2006.
\BIBentrySTDinterwordspacing

\bibitem{Mairal2010}
J.~Mairal, F.~Bach, J.~Ponce, and G.~Sapiro, ``{Online Learning for Matrix
  Factorization and Sparse Coding},'' \emph{The Journal of Machine Learning
  Research}, vol.~11, pp. 19--60, 2010.

\bibitem{Scutari2012a}
\BIBentryALTinterwordspacing
G.~Scutari, F.~Facchinei, J.-S. Pang, and D.~P. Palomar, ``{Real and Complex
  Monotone Communication Games},'' \emph{IEEE Transactions on Information
  Theory}, vol.~60, no.~7, pp. 4197--4231, Jul. 2014.
\BIBentrySTDinterwordspacing

\bibitem{Hjorungnes}
A.~Hj{\o}rungnes, \emph{{Complex-valued matrix derivatives with applications in
  signal processing and communications}}.\hskip 1em plus 0.5em minus
  0.4em\relax Cambridge: Cambridge University Press, 2011.

\bibitem{Zhang2008b}
\BIBentryALTinterwordspacing
J.~Zhang, D.~Zheng, and M.~Chiang, ``{The Impact of Stochastic Noisy Feedback
  on Distributed Network Utility Maximization},'' \emph{IEEE Transactions on
  Information Theory}, vol.~54, no.~2, pp. 645--665, Feb. 2008.
\BIBentrySTDinterwordspacing

\bibitem{Hong2011}
\BIBentryALTinterwordspacing
M.~Hong and A.~Garcia, ``{Averaged Iterative Water-Filling Algorithm:
  Robustness and Convergence},'' \emph{IEEE Transactions on Signal Processing},
  vol.~59, no.~5, pp. 2448--2454, May 2011.
\BIBentrySTDinterwordspacing

\bibitem{DiLorenzo2013}
\BIBentryALTinterwordspacing
P.~{Di Lorenzo}, S.~Barbarossa, and M.~Omilipo, ``{Distributed Sum-Rate
  Maximization Over Finite Rate Coordination Links Affected by Random
  Failures},'' \emph{IEEE Transactions on Signal Processing}, vol.~61, no.~3,
  pp. 648--660, Feb. 2013.
\BIBentrySTDinterwordspacing

\bibitem{Huang2006a}
\BIBentryALTinterwordspacing
J.~Huang, R.~A. Berry, and M.~L. Honig, ``{Distributed interference
  compensation for wireless networks},'' \emph{IEEE Journal on Selected Areas
  in Communications}, vol.~24, no.~5, pp. 1074--1084, 2006.
\BIBentrySTDinterwordspacing

\bibitem{Shi2009a}
\BIBentryALTinterwordspacing
C.~Shi, R.~A. Berry, and M.~L. Honig, ``{Monotonic convergence of distributed
  interference pricing in wireless networks},'' in \emph{2009 IEEE
  International Symposium on Information Theory}.\hskip 1em plus 0.5em minus
  0.4em\relax IEEE, Jun. 2009, pp. 1619--1623.
\BIBentrySTDinterwordspacing

\bibitem{Kim2011}
\BIBentryALTinterwordspacing
S.-J. Kim and G.~B. Giannakis, ``{Optimal Resource Allocation for MIMO Ad Hoc
  Cognitive Radio Networks},'' \emph{IEEE Transactions on Information Theory},
  vol.~57, no.~5, pp. 3117--3131, May 2011.
\BIBentrySTDinterwordspacing

\bibitem{Luo2008}
\BIBentryALTinterwordspacing
Z.-Q. Luo and S.~Zhang, ``{Dynamic Spectrum Management: Complexity and
  Duality},'' \emph{IEEE Journal of Selected Topics in Signal Processing},
  vol.~2, no.~1, pp. 57--73, Feb. 2008.
\BIBentrySTDinterwordspacing

\bibitem{Bertsekas}
D.~P. Bertsekas and J.~N. Tsitsiklis, \emph{{Parallel and distributed
  computation: Numerical methods}}.\hskip 1em plus 0.5em minus 0.4em\relax
  Prentice Hall, 1989.

\bibitem{Ram2009}
\BIBentryALTinterwordspacing
S.~{Sundhar Ram}, A.~Nedi\'{c}, and V.~V. Veeravalli, ``{Incremental Stochastic
  Subgradient Algorithms for Convex Optimization},'' \emph{SIAM Journal on
  Optimization}, vol.~20, no.~2, pp. 691--717, Jan. 2009.
\BIBentrySTDinterwordspacing

\bibitem{Srivastava2011}
\BIBentryALTinterwordspacing
K.~Srivastava and A.~Nedi\'{c}, ``{Distributed Asynchronous Constrained
  Stochastic Optimization},'' \emph{IEEE Journal of Selected Topics in Signal
  Processing}, vol.~5, no.~4, pp. 772--790, Aug. 2011.
\BIBentrySTDinterwordspacing

\bibitem{bertsekas03}
D.~P. Bertsekas, A.~Nedic, and A.~E. Ozdaglar, \emph{{Convex Analysis and
  Optimization}}.\hskip 1em plus 0.5em minus 0.4em\relax Athena Scientific,
  2003.

\bibitem{bertsekas1999nonlinear}
D.~P. Bertsekas, \emph{{Nonlinear programming}}.\hskip 1em plus 0.5em minus
  0.4em\relax Athena Scientific, 1999.

\bibitem{Razaviyayn_spawc2013}
M.~Razaviyayn, M.~Sanjabi, and Z.-Q. Luo, ``{A stochastic weighted MMSE
  approach to sum rate maximization for a MIMO interference channel},'' in
  \emph{The 14th IEEE International Workshop on Signal Processing Advances for
  Wireless Communications}, 2013.

\bibitem{boyd2004convex}
S.~Boyd and L.~Vandenberghe, \emph{{Convex optimization}}.\hskip 1em plus 0.5em
  minus 0.4em\relax Cambridge Univ Pr, 2004.

\bibitem{Rockafellar70}
R.~T. Rockafellar, \emph{{Convex Analysis}}, 2nd~ed.\hskip 1em plus 0.5em minus
  0.4em\relax Princeton, NJ: Princeton Univ. Press, 1970.

\bibitem{Yu2004}
\BIBentryALTinterwordspacing
W.~Yu, W.~Rhee, S.~Boyd, and J.~Cioffi, ``{Iterative Water-Filling for Gaussian
  Vector Multiple-Access Channels},'' \emph{IEEE Transactions on Information
  Theory}, vol.~50, no.~1, pp. 145--152, Jan. 2004.
\BIBentrySTDinterwordspacing

\end{thebibliography}

\end{document}